\documentclass{PoS}

\usepackage{latexsym}
\usepackage{graphicx}
\usepackage{epstopdf}
\usepackage{bm}
\title{Nucleon structure with dynamical (2+1)-flavor domain wall fermions lattice QCD}

\ShortTitle{(2+1)-flavor dynamical dwf nucleon structure}

\author{Shigemi Ohta\thanks{Speaker of the structure function part.}\\
        Institute of Particle and Nuclear Studies, KEK, Tsukuba, Ibaraki 305-0801, Japan\\
        RIKEN-BNL Research Center, Brookhaven National Laboratory, Upton, NY 11973\\
        Physics Department, Sokendai Grad.\ Univ.\ Adv.\ Studies, Hayama, Kanagawa 240-0193, Japan\\
        E-mail: \email{shigemi.ohta@kek.jp}}

\author{Takeshi Yamazaki\thanks{Speaker of the form factor part.}\\
        Yukawa Institute for Theoretical Physics , Kyoto University, Kyoto 606-8502, Japan\\
        E-mail: \email{tyamazak@yukawa.kyoto-u.ac.jp}}
        
\author{RBC and UKQCD Collaborations}

\abstract{
We report isovector form factors and low moments of isovector structure functions of nucleon from the coarse RIKEN-BNL-Columbia (RBC) and UKQCD joint dynamical (2+1)-flavor domain-wall fermions (DWF) ensembles.
The lattice cut off is estimated at \(a^{-1}=1.7\) GeV.
The lattice volume is as large as 2.7 fm across.
The dynamical strange mass is set at slightly heavier than physical, and degenerate up and down mass is varied correpsonding to pion mass of about 0.67, 0.56, 0.42 and 0.33 GeV.
We carefully optimize the nucleon source/sink separation in time to about 1.4 fm.
Unexpectedly large finite-size effect in the axial charge is found.
The effect scales with a single variable, the product \(m_\pi L\) of the pion mass \(m_\pi\) and lattice spatial linear extent \(L\), and sets in at around \(m_\pi L = 5\).
We also discuss momentum-transfer dependence of the vector, induced tensor, axial-vector and induced pesudo-scalar form factors.
In particular we extract Dirac and Pauli mean-squared charge radii, magnetic moment, \(\pi NN\) coupling constant and pseudoscalar coupling \(g_P\) of muon capture by nucleon.
While the mean-squared charge radii are distant from the respective experiments, the other observables are in reasonable agreement with the experiments.
From structure functions, fully non-perturbatively renormalized iso-vector quark momentum fraction, \(\langle x \rangle_{u-d}\), helicity fraction, \(\langle x \rangle_{\Delta u - \Delta d}\), and transversity, \(\langle 1 \rangle_{\delta u - \delta d}\), are reported, as well as an unrenormalized twist-3 coefficient, \(d_1\).
The ratio of the momentum to helicity fractions, \(\langle x \rangle_{u-d}/\langle x \rangle_{\Delta u - \Delta d}\), does not depend on light quark mass and agree well with the experiment.
Their respective absolute values, fully renormalized, shows interesting trending toward the respective experimental values at the lightest light quark mass.

\vspace{-211mm}\parbox{\textwidth}{\flushright\large\rm \hfill KEK-TH-1232, RBRC-756, YITP-08-74}\vspace{211mm}
}

\FullConference{The XXVI International Symposium on Lattice Field Theory\\
		 July 14-19 2008\\
		 Williamsburg, Virginia, USA}

\begin{document}

\section{Introduction}

We report numerical lattice quantum chromodynamics (QCD) calculations of all the four isovector form factors and some low moments of the isovector structure functions of nucleon using the lattice gauge ensembles \cite{Allton:2008pn} jointly generated by the RIKEN-BNL-Columbia (RBC) and UKQCD Collaborations with ``2+1'' flavors of dynamical domain-wall fermions (DWF)~\cite{Kaplan:1992bt,Shamir:1993zy,Furman:1994ky}.

Four isovector form factors parameterize neutron $\beta$ decay: the vector and induced tensor form factors from the vector current,
\begin{equation}\label{eq:cont_vector}
\langle p| V^+_\mu(x) | n \rangle = \bar{u}_p \left[\gamma_\mu
G_V(q^2) -q_{\lambda} \sigma_{\lambda \mu} {G_T(q^2)} \right]
u_n e^{iq\cdot x},
\end{equation}
where \(G_V\) is equivalent to \(F_1\) and \(G_T\) to \(F_2/(2m_N)\) in electromagnetic form factors,
and the axial and induced pseudoscalar form factors from the axial current,
\begin{equation}\label{eq:cont_axial}
\langle p| A^+_\mu(x) | n \rangle = \bar{u}_p
            \left[\gamma_\mu  \gamma_5 G_{A}(q^2)
             +i q_\mu \gamma_5 {G_{P}(q^2)} \right]  u_n e^{iq\cdot x}.
\end{equation}
We use the Euclidean metric convention: thus $q^2$ stands for Euclidean four-momentum squared, and corresponds to the time-like momentum squared as $q_M^2=-q^2<0$ in Minkowski space.
Here $q=p_n-p_p$ is the momentum transfer between the proton ($p$) and neutron ($n$).
In the limit $|{\vec q}| \rightarrow 0$, the momentum transfer should be small because the mass difference of the neutron and proton is only about 1.3~MeV.
This makes the limit $q^2 \rightarrow 0$, where the vector and axial form factors dominate, a good approximation.
Their values in this limit are called the vector and axial charges of the nucleon: $g_V = G_V(q^2=0)$ and $g_A = G_A(q^2=0)$.
As is well known, the vector charge is determined by the Cabibbo mixing angle $\theta_C$: $g_V = \cos \theta_C$.
The axial charge receives strong interaction correction, and is very accurately known, $g_A = 1.2695(29) \times g_V$~\cite{Yao:2006px} from neutron \(\beta\) decay measurements.
It is an interesting and important challenge for numerical lattice QCD to reproduce this value.

Nucleon form factor experiments of course are not limited to neutron \(\beta\) decays.
Vast data of lepton elastic scattering exist at wide range of momentum transfer values.
On the lattice we can investigate nucleon form factors at relatively low momentum transfer such as \(\vec{q} = \sim 2\pi/L \times (0,0,0)\), (0,0,1), (0,1,1), (1,1,1), (0,0,2), characterized by the inverse of the linear lattice spatial extent \(L\).
From the dependence on momentum-trasfer we can extract mean-squared radii of the respective form factors, anomalous magnetic moment, \(\pi NN\) coupling, pseudoscalar effective coupling \(g_P\) that enters the muon capture by nucleon, and so on.

The structure functions are measured in deep inelastic scattering of leptons from nucleons, the cross section of which is factorized in terms of leptonic and hadronic tensors,
\(
\propto
l^{\mu\nu}W_{\mu\nu}.
\)
Since the leptonic tensor, \(l_{\mu\nu}\), is known, the cross section provides us with structure information about the target nucleon through the hadronic tensor, $W_{\mu\nu}$, which is decomposed into symmetric unpolarized and antisymmetric polarized parts:
\begin{equation}\label{eq:Wsym}
W^{\{\mu\nu\}}(x,Q^2) =
\left( -g^{\mu\nu} + \frac{q^\mu
q^\nu}{q^2}\right)  {F_1(x,Q^2)} +
\left(P^\mu-\frac{\nu}{q^2}q^\mu\right)\left(P^\nu-\frac{\nu}{q^2}q^\nu\right)
\frac{F_2(x,Q^2)}{\nu},
\end{equation}
\begin{equation}\label{eq:Want}
W^{[\mu\nu]}(x,Q^2) = i\epsilon^{\mu\nu\rho\sigma} q_\rho
\left(\frac{S_\sigma}{\nu}({g_1(x,Q^2)} \right. +
 \left. {g_2(x,Q^2)}) - \frac{q\cdot S
P_\sigma}{\nu^2}{g_2(x,Q^2)} \right),
\end{equation}
with kinematic variables defined as $\nu = q\cdot P$, $S^2 = -M^2$, and $x=Q^2/2\nu$, and \(Q^2=|q^2|\).
The unpolarized structure functions are $F_1(x,Q^2)$ and $F_2(x,Q^2)$, and the polarized, $g_1(x,Q^2)$ and $g_2(x,Q^2)$.
Their moments are described in terms of Wilson's operator product expansion:
\begin{eqnarray}\label{eq:moments}
2 \int_0^1 dx\,x^{n-1} {F_1(x,Q^2)} &=& \sum_{q=u,d}
c^{(q)}_{1,n}\: \langle x^n \rangle_{q}(\mu)
+{O(1/Q^2)},
\nonumber \\
\int_0^1 dx\,x^{n-2} {F_2(x,Q^2)} &=& \sum_{f=u,d}
c^{(q)}_{2,n}\: \langle x^n \rangle_{q}(\mu)
+{O(1/Q^2)},
\nonumber \\
2\int_0^1 dx\,x^n {g_1(x,Q^2)}
  &=& \sum_{q=u,d} e^{(q)}_{1,n}\: \langle x^n \rangle_{\Delta q}(\mu)
+{O(1/Q^2)},  \nonumber
 \\
2\int_0^1 dx\,x^n {g_2(x,Q^2)}
  &=& \frac{1}{2}\frac{n}{n+1} \sum_{q=u,d} \left[e^{q}_{2,n}\: d_n^{q}(\mu)
-  2 e^{q}_{1,n}\: \langle x^n \rangle_{\Delta
q}(\mu)\right] + {O(1/Q^2)},
\end{eqnarray}
where we suppressed the \((\mu^2/Q^2,g(\mu))\) dependence of the perturbatively known Wilson coefficients, $c_1$, $c_2$, $e_1$, and $e_2$.
The moments, ${\langle x^n \rangle_{q}(\mu)}$, ${\langle x^n \rangle_{\Delta q}(\mu)}$ and $d_n(\mu)$ are calculable on the lattice as forward nucleon matrix elements of certain local operators.

In addition,  the tensor charge, \(\langle 1\rangle_{\delta q}(\mu)\), which probes transverse spin structure of nucleon, is beginning to be reported by experiments \cite{Anselmino:2007fs,RHICSpin}.
This quantity is also calculable on the lattice much like the way the DIS structure function moments are calculated.

In the following we report our lattice numerical calculations of the four iso-vector form factors, \(G_V(q^2)\), \(G_T(q^2)\), \(G_A(q^2)\), and \(G_P(q^2)\), and four iso-vector moments of the structure functions, the quark momentum fraction, \(\langle x\rangle_{u-d}(\mu)\), the helicity fraction, \(\langle x\rangle_{\Delta u-\Delta d}(\mu)\), the tensor charge, \(\langle 1\rangle_{\delta u - \delta d}(\mu)\), and twist-3 coefficient \(d_1\) of the \(g_2\) polarized structure function.
All but \(d_1\) are fully non-perturbatively renormalized.

\section{Formulation}

We use the standard proton operator, \(B = \epsilon_{abc} (u_a^T C \gamma_5d_b)u_c\) to create and annihilate proton states.
We Gaussian-smear this operator for better overlap with the ground state with both zero and finite momentum.
A Gaussian radius of 7 is chosen after a series of pilot calculations.
Since the up and down quark mass are degenerate in these calculations, the isospin symmetry is exact.
Thus we often take advantage of rotation in isospin space through Wigner-Eckhart theorem from proton-neutron matrix elements, \(\langle {\rm neutron} | O | {\rm proton}\rangle\), to proton-proton ones, \(\langle {\rm proton} | O | {\rm proton}\rangle\).
This is of course a well-known good approximation.
We project the positive-parity ground state, so our two-point proton function takes the form
\begin{equation}
C_{\rm 2pt}(t) =
\sum_{\alpha,\beta}
\left(\frac{1+\gamma_t}{2}\right)_{\alpha\beta} 
\langle B_\beta(t_{\rm sink})\overline{B}_\alpha(t_{\rm source}) \rangle,
\end{equation}
with \(\displaystyle  t=t_{\rm sink}-t_{\rm source}\).
We insert an appropriate observable operator \(O(\vec{q}, \tau)\) at time \(\tau\), \(t_{\rm source}\le \tau \le t_{\rm sink}\), and possibly finite momentum transfer \(\vec{q}\), to obtain a form factor or structure function moment three-point function,
\begin{equation}
C^{\Gamma, O}_{\rm 3pt}(t, \tau. \vec{q}) =
\sum_{\alpha,\beta}
\Gamma_{\alpha\beta}
\langle B_\beta(t_{\rm sink}) O(\vec{q}, t) \overline{B}_\alpha(t_{\rm source}) \rangle,
\end{equation}
with appropriate projection, \(\displaystyle \Gamma=\frac{1+\gamma_t}{2}\), for the spin-unpolarized, and \(\displaystyle \Gamma=\frac{1+\gamma_t}{2}i\gamma_5\gamma_k, k\ne 4\), for the polarized.
Ratios of these two- and three-point functions give plateaux for \(0<\tau<t\) that give the bare lattice matrix elements of desired observables: e.g.\ 
\(\displaystyle
\langle O\rangle^{\rm bare}=
\frac{C^{\Gamma, O}_{\rm 3pt}(t, \tau) }{C_{\rm 2pt}(t)}
\)
at \(q^2=0\).
At finite \(q^2\) we need to take care of extra kinematics.
We renormalize structure function moments by Rome-Southampton RI-MOM non-perturbative renomalization prescription \cite{Dawson:1997ic}.
The good continuum-like flavor and chiral symmetries of domain-wall fermions are very useful here in eliminating unwanted lattice-artifact mixings that are present in many other fermion schemes.
For form factors this provides natural lattice renormalization through the vector charge.
See our earlier publications, ref.\ \cite{Lin:2008uz}, and references cited there in for further details.

The RBC-UKQCD joint (2+1)-flavor dynamical DWF coarse ensembles \cite{Allton:2008pn} are used for the calculations.
These ensembles are generated with Iwasaki action at the coupling \(\beta=2.13\) which corresponds to
the lattice cut off of about \(a^{-1}=1.73(3)\) GeV.
There are two lattice volumes, \(16^3\times 32\) and \(24^3\times 64\), corresponding to linear spatial extent of  about 1.8 and 2.7 fm.
The dynamical strange and up and down quarks are described by DWF actions with the fifth-dimensional mass of \(M_5=1.8\).
The strange mass is set at 0.04 in lattice unit and turned out to be about twelve percent heavier than physical including the additive correction of the residual mass, \(m_{\rm res}=0.003\).
The degenerate up and down mass is varied at 0.03, 0.02, 0.01 and 0.005, correpsonding to pion mass of about 0.67, 0.56, 0.42 and 0.33 GeV and nucleon mass are about 1.55, 1.39, 1.22 and 1.15 GeV.
We summarize the gauge configurations and pion and nucleon mass of each ensemble in Table \ref{table:statistics}.
\begin{table}[t]
\begin{center}
\begin{tabular}{lccccc} \hline
\multicolumn{1}{c}{$m_fa$} & 
\multicolumn{1}{c}{\# of config.'s} &
\multicolumn{1}{c}{meas.\ interval} &
\multicolumn{1}{c}{$N_{\rm sources}$} &
\multicolumn{1}{c}{$m_\pi$ (GeV)} &
\multicolumn{1}{c}{$m_N$ (GeV)}\\ \hline
0.005 & 932 & 10 & 4 & 0.33 &1.15\\ 
0.01  & 356 & 10 & 4 & 0.42 &1.22\\
0.02  &  98 & 20 & 4 & 0.56 &1.39\\ 
0.03  & 106 & 20 & 4 & 0.67 &1.55\\ \hline
\end{tabular}
\end{center}
\caption{Number of gauge configurations and pion and nucleon mass.\label{table:statistics}}
\end{table}

There are two important sources of systematic error: finite spatial size of the lattice and excited states contamination:
Chiral-perturbation-inspired analysis of the former for meson observables suggests the dimensionless product, \(m_\pi L\), of the calculated pion mass, \(m_\pi\) and lattice linear spatial extent, \(L\), should be set greater than 4 to drive the finite-volume correction below one percent, and the available lattice calculations seem to support this.
While our present parameters satisfy this condition, it should be emphasized that this criterion is not known sufficient for baryon observables.
It is important to check this through the present calculations, and it is indeed an important purpose.

One should adjust the time separation, \(t\), between the nucleon source and sink appropriately so the resultant nucleon observables are free of contamination from excited states.
The separation has to be made longer as we set the quark masses lighter.
Here our previous study with two dynamical flavors of DWF quarks \cite{Lin:2008uz} help as the lattice cutoff is similar at 1.7 GeV.
\begin{figure}
\begin{center}
\includegraphics[width=.48\textwidth,clip]{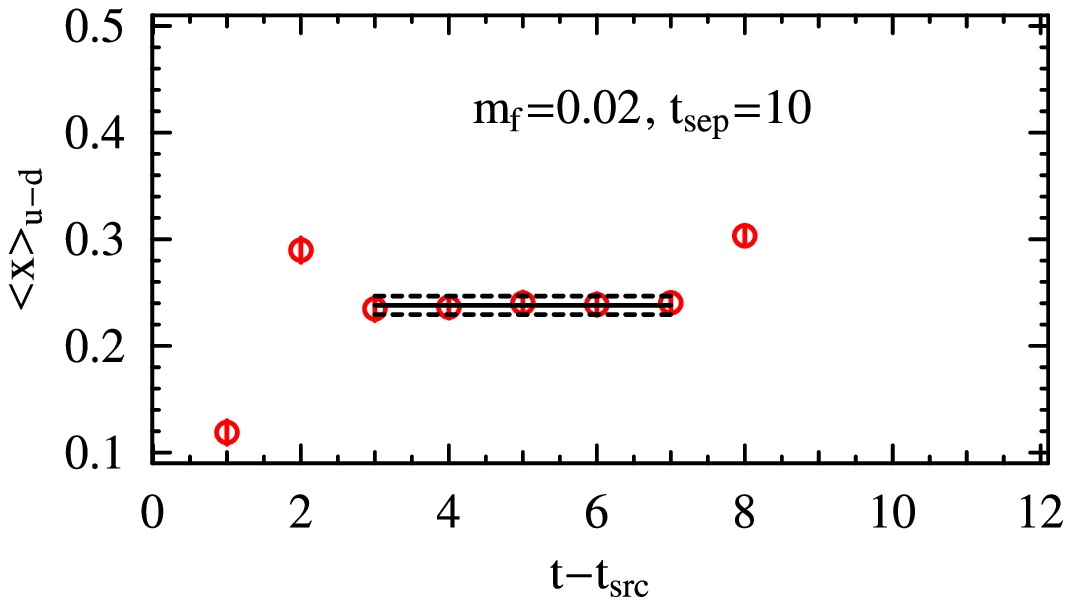}
\includegraphics[width=.48\textwidth,clip]{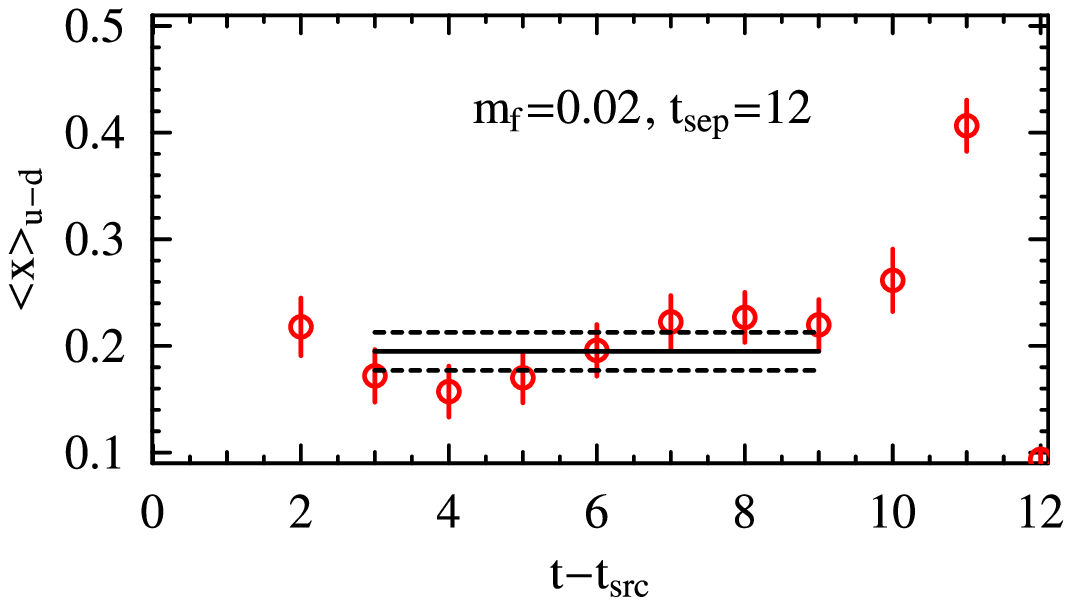}
\end{center}
\caption{A nucleon observable, isovector quark momentum fraction, \(\langle x\rangle_{u-d}\), from RBC 2-flavor dynamical DWF ensemble with \(m_{ud}a=0.02\) \cite{Lin:2008uz}, with source-sink separation of 10 and 12: a clear systematic difference is seen.  The shorter source-sink separation is not manifestly free of excited-state contamination.\label{fig:source-sink}}
\end{figure}
In Figure \ref{fig:source-sink} we compare a nucleon observable, isovector quark momentum fraction, \(\langle x\rangle_{u-d}\), at the up/down mass of 0.02 where pion mass is about 500 MeV and nucleon 1.3 GeV.
A clear systematic difference is seen between the shorter time separation of 10, or about 1.16 fm, and longer 12, or 1.39 fm: the former averages about 0.24 while the latter about 0.20.
Although the former appears statistically better, as the two-point function that provides the normalization has not decayed much yet, it is not manifestly free of excited-state contamination that is expected more at shorter time separation.
The latter suffers statistically, as the two-point function that provides the normalization has decayed, but is certainly freer of excited-state contamination than the former: Of the two, we have to choose the latter.
If larger statistics is required, then we have to collect as much statistics as necessary for the latter.
Ideally we should collect more statistics at even larger source-sink separation to confirm the excited-state contamination has been eliminated by the separation of 12.

Since in the present ensembles we explore much lighter up/down quark mass than this two-flavor example, with pion mass as light as 330 MeV and nucleon 1.15 GeV, the source/sink time separation must be chosen longer: longer than 1.2 fm at least.
\begin{figure}[b]
\begin{center}
\includegraphics[width=\textwidth,clip]{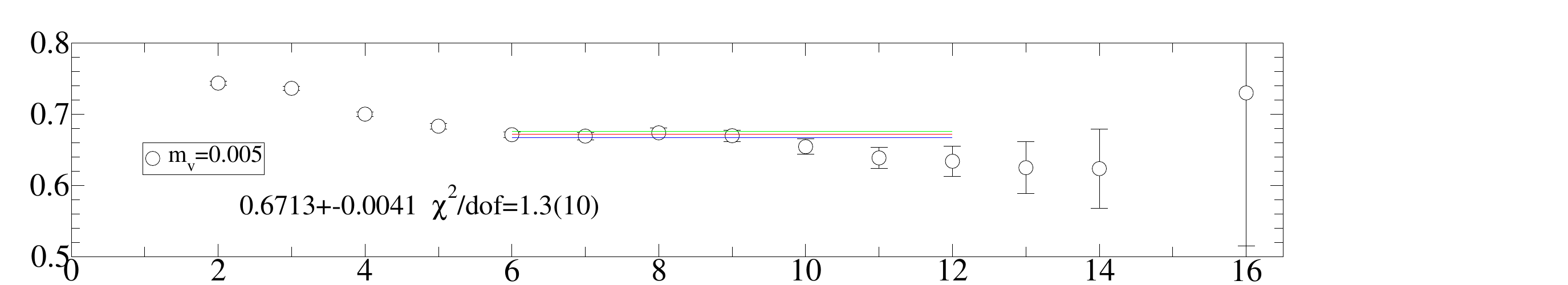}
\end{center}
\caption{Nucleon effective mass at quark mass \(m_fa=0.005\) in the present calculation with (2+1)-flavor dynamical DWF quarks.  The nucleon signal begins to decay at time separation 12.  By which time we are as free of excited state as possible while maintaining the nucleon ground state.\label{fig:Nemass}}
\end{figure}
\begin{figure}
\begin{center}
\includegraphics[width=.48\textwidth,clip]{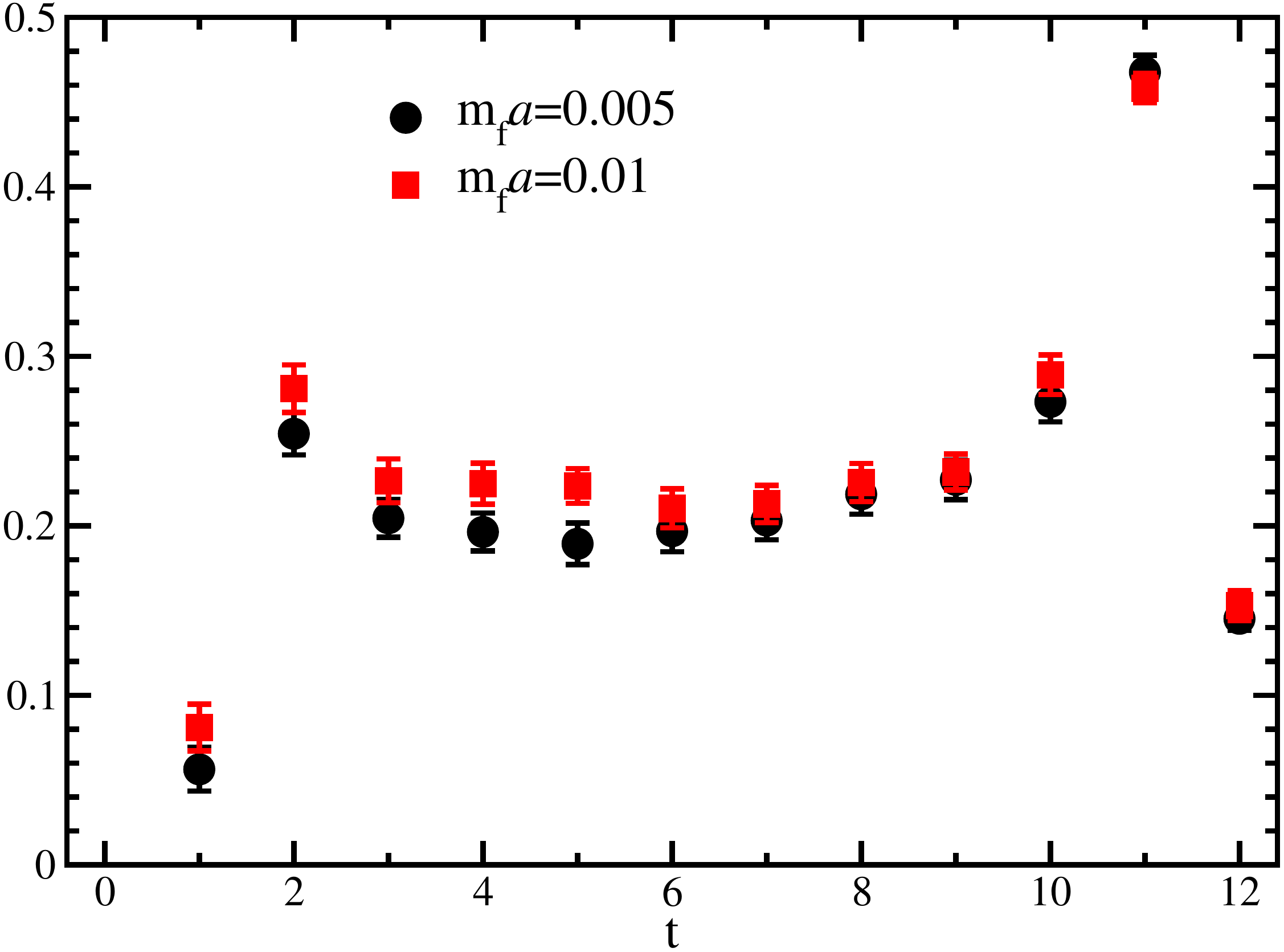}
\includegraphics[width=.48\textwidth,clip]{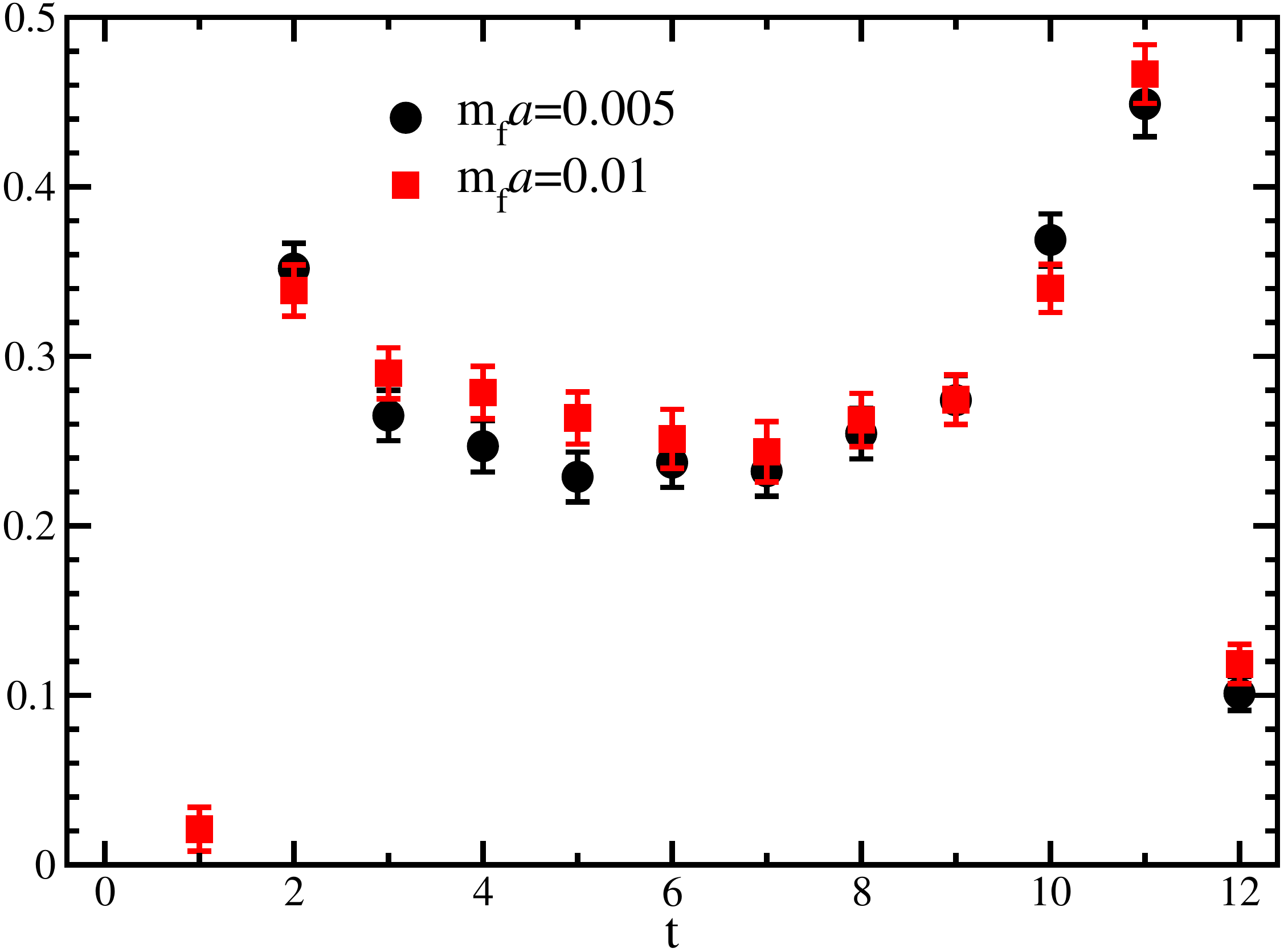}
\end{center}
\caption{Bare three-point function for quark momentum fraction, \(\langle x\rangle_{u-d}\), (left) and helicity fraction, \(\langle x\rangle_{\Delta u - \Delta d}\), (right) for \(m_fa=0.005\) and 0.01.\label{fig:signals}}
\end{figure}
In Figure \ref{fig:Nemass} we present the nucleon effective mass at the lightest up/down quark mass, \(m_fa=0.005\), in the present (2+1)-flavor dynamical DWF calculation.
The nucleon signal begins to decay at \(t=12\), or about 1.4 fm: this is about longest distance we can choose without losing the signal, and hence about as free of excited-state contamination as we can go. As is shown in Figure \ref{fig:signals} the bare three-point function signals for quark momentum and helicity fractions for this source-sink separation of \(t=12\) are acceptable.
Thus we choose the source-sink separation of 12, or about 1.4 fm for the present work, as the two-flavor example suggests strongly that anything shorter would suffer from excited-state contamination.
To confirm this choice is sufficiently long to eliminate the excited-state contamination, we need to collect more statistics with a longer source-sink separation.

\section{Axial charge}

This part has been published \cite{Yamazaki:2008py}: here we repeat the essence for the readers' convenience.
\begin{figure}[b]
\label{fig:gagv}
\begin{center}
\includegraphics[width=.48\textwidth,clip]{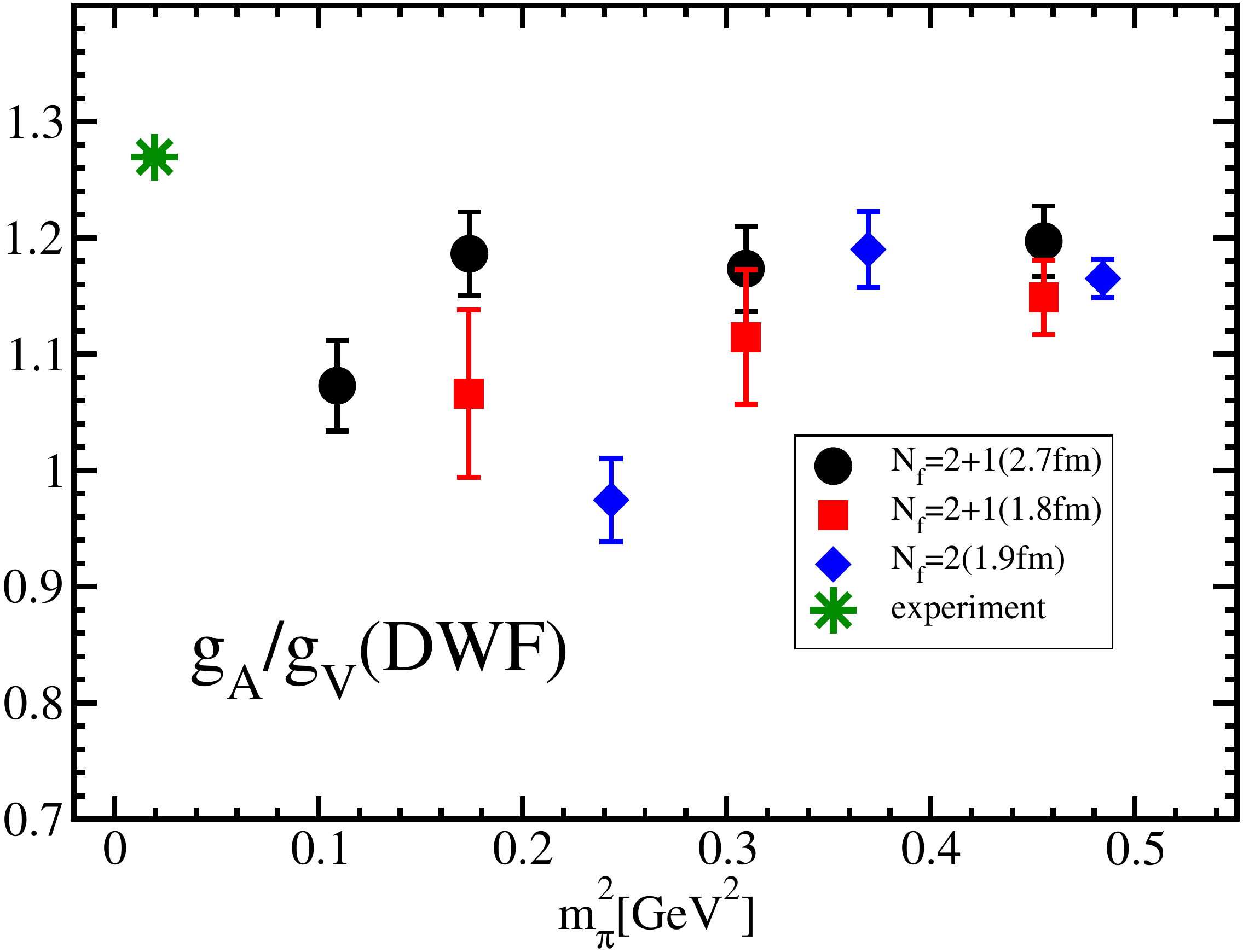}
\includegraphics[width=.48\textwidth,clip]{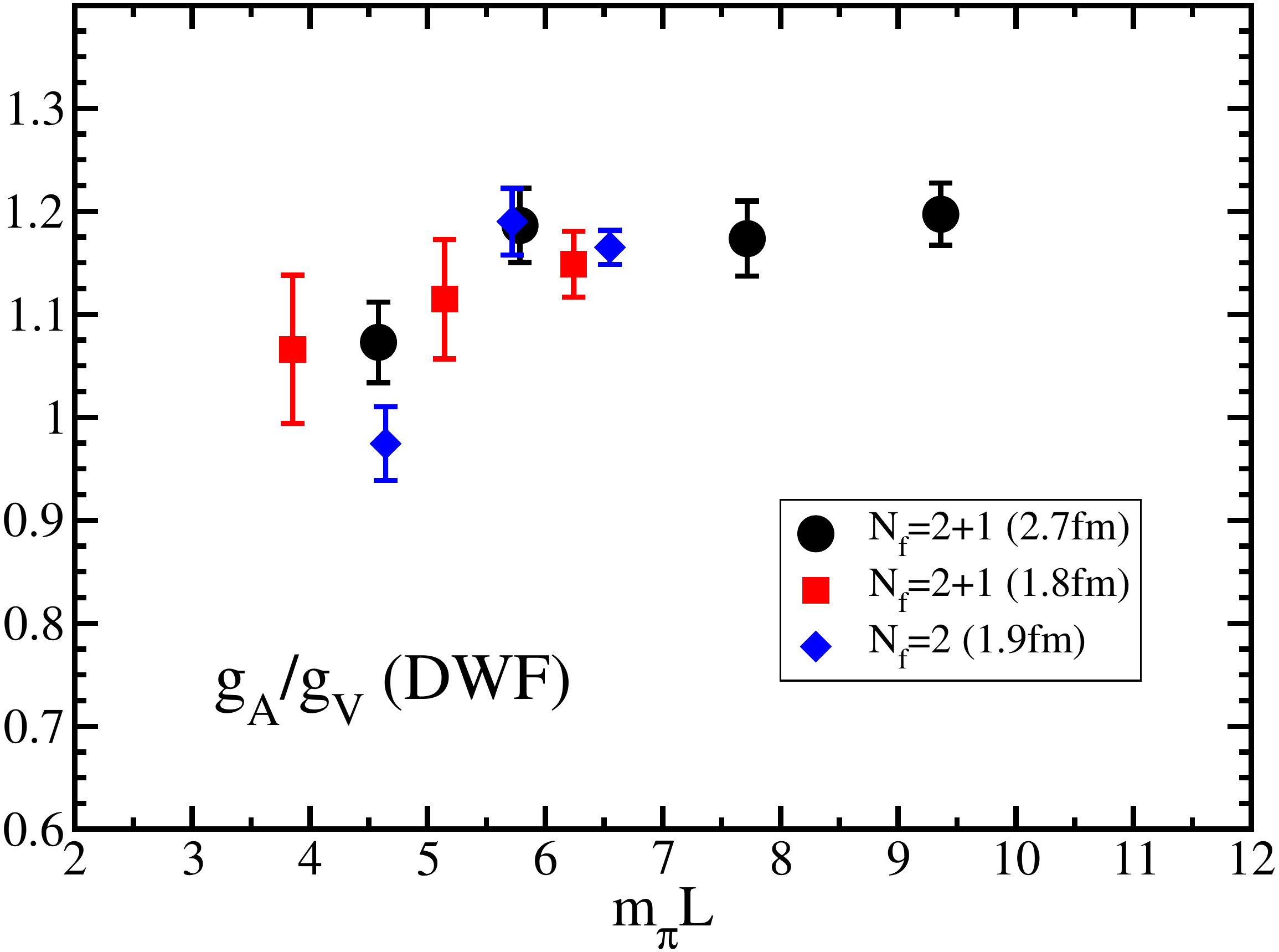}
\end{center}
\caption{Left pane: Lattice ratio, \(g_A/g_V\), plotted against pion mass squared, \(m_\pi^2\) for the current RBC/UKQCD (2+1)-flavor ensembles, large (\(24^3=({\rm 2.7 fm})^3\), circles) and small (\(16^3=({\rm 1.8 fm})^3\), squares) lattice, and old RBC 2-flavor (diamonds) ensembles.}
\end{figure}
The isovector axial charge is the value of the axial form factor at zero four-momentum transfer, \(g_A = G_A(0)\).
Since the DWF action preserves the chiral symmetry, the quark vector and axial currents share a common renormalization up to higher-order deiscretization errors, \(O((m_fa)^2)\).
Hence the lattice ratio, \((g_A/g_V)^{\rm lattice}\), to the vector charge, is already renormalized and is directly comparable with the experiment.
In the left pane of Figure \ref{fig:gagv} we summarize the results for this ratio from the current RBC/UKQCD (2+1)-flavor dynamical DWF ensembles on large (2.7 fm) and small (1.8 fm) volumes, all plotted against the calculated pion mass squared, \(m_\pi^2\).

In the larger, 2.7 fm, volume of the present calculation, the three heavier quark mass points at \(m_fa=0.01\), 0.02 and 0.03 do not show much dependence on quark mass and are close to the experimental value.
However at the lightest quark mass, \(m_fa=0.005\), the value deviates significantly downward and away from the experiment.
In the smaller, 1.8 fm, volume, the central values all deviate downward from the corresponding large-volume results, albeit with large statistical errors.
This suggests the downward shift is caused by the small finite spatial volumes of the lattice.
It is likely such a finite-size effect is enhanced at lighter quark mass. 

The same ratio in the old RBC 2-flavor dynamical DWF calculation at a similar cutoff and with a lattice size comparable to the current smaller volume, 1.9 fm across, show similar trend that the heavier two point stay close with each other and with the experiment, and the lightest point deviating downward away from the experiment.
These small-volume results suggest the downward shift away from the experiment may be caused by the finite-size effect.
Indeed if we plot these values against another variable, the dimensionless product, \(m_\pi L\), of the calculated pion mass, \(m_\pi\), and the lattice linear spatial extent, \(L\), then the points align on a monotonically increasing curve, as is shown in the left pane of Figure \ref{fig:gagv}.
In other words, the quantity scales with the variable \(m_\pi L\).
Thus we conclude the downward shifts are finite-size effect.

We also note that an earlier quenched RBC calculation \cite{Sasaki:2003jh} saw similar finite-size effect.
However in these quenched calculations the effect was not detected till the linear spatial extent of the lattice was made shorter than 2.4 fm.
It appears that the finite-size effect is enhanced by dynamical quarks.

A similar scaling in this variable, \(m_\pi L\), is obtained \cite{Yamazaki:2008py} if we replot 2-flavor Wilson fermion calculations of the same quantity \cite{Khan:2006de,Dolgov:2002zm,Alexandrou:2007zz}, corroborating our discovery of the \(m_\pi L\)-scaling here.
\begin{figure}
\begin{center}
\includegraphics[width=.48\textwidth,clip]{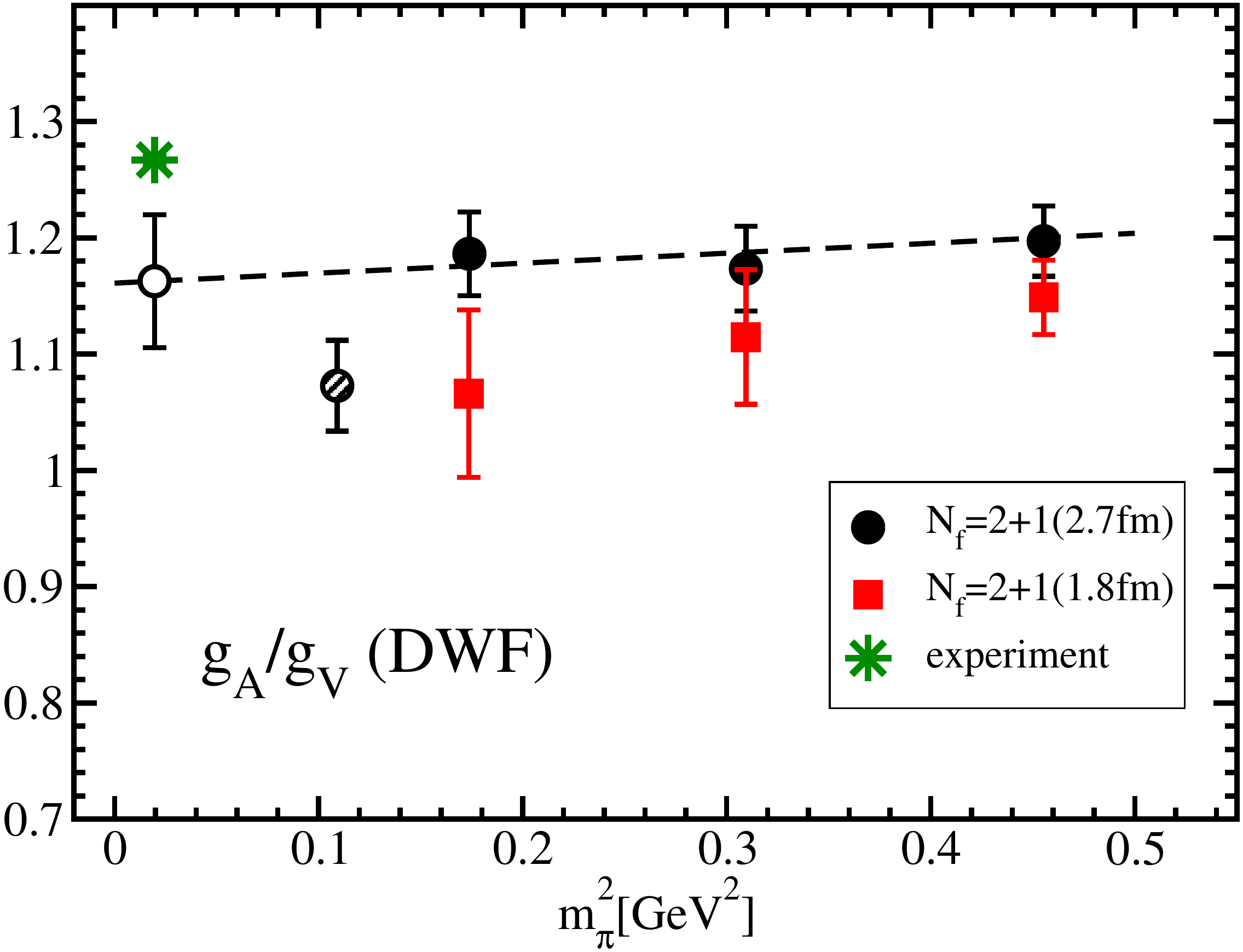}
\includegraphics[width=.48\textwidth,clip]{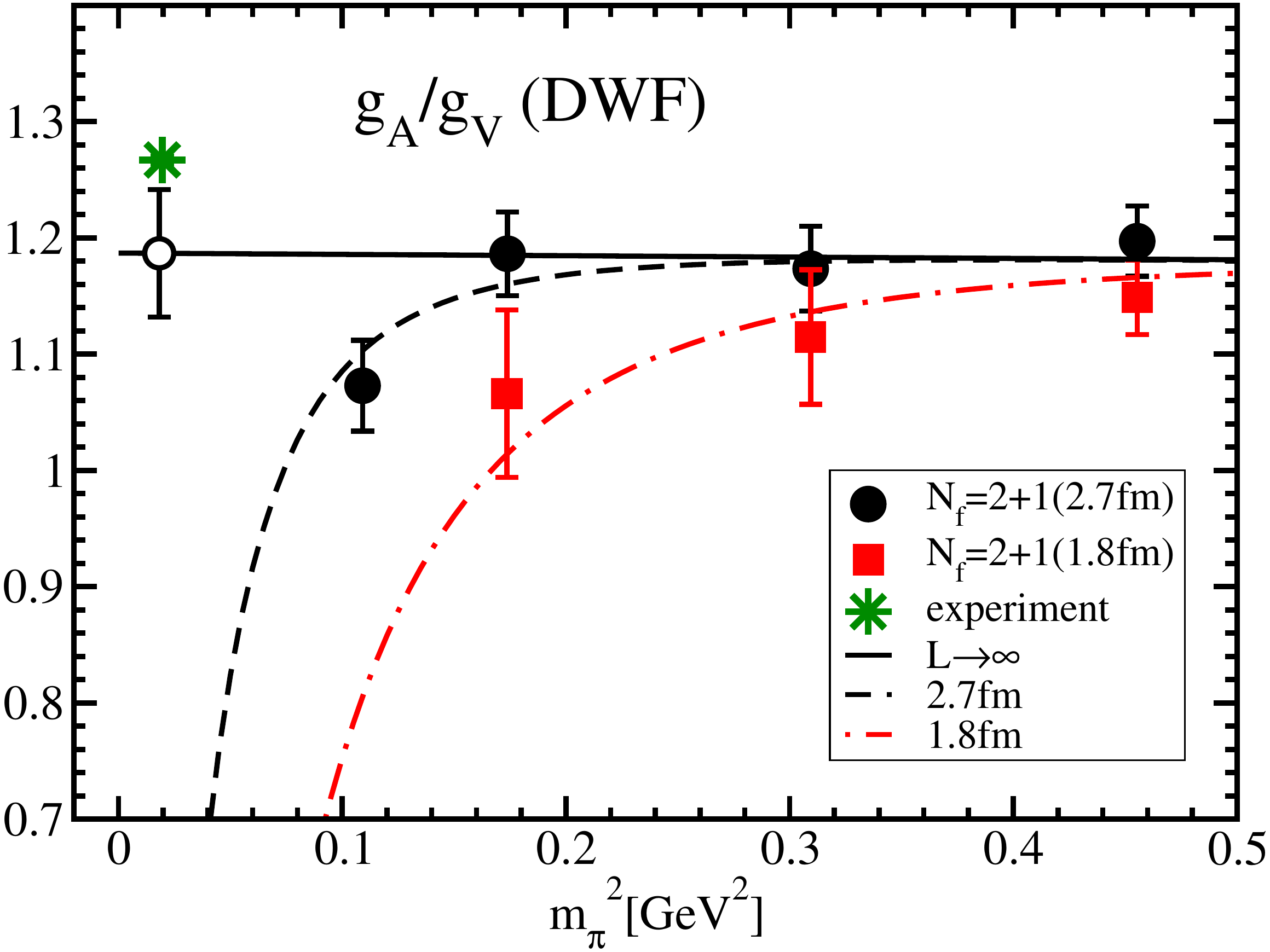}
\end{center}
\caption{Left: Linear fit to the three heaviest points from the larger, 2.7 fm, volume, of calculated axial charge, \(g_A/g_V\).  Right: an example of simultaneous fit with finite-size effect to all the results from both large and small volume results.\label{fig:gafits}}
\end{figure}

In light of this unexpectedly large finite-size effect, apparently due to light dynamical quarks, we need much larger volumes to accurately calculate the axial charge, at least, and likely other elastic form factors.
Since the finite-size effect grows larger than 1 \% at \(m_\pi L \sim 5\), we would need a 5 fm across lattice for moderately light 200 MeV pions and 7 fm for realistic pions.

On the other hand we can exploit the finite-size effect in extrapolating to the physical pion mass.
The simplest way is to drop all the results where finite-size effect is seen and use only the three heaviest points in the large volume which comfortably accommodate a linear fit: we obtain a value 1.17(6) for \(g_A/g_V\) at physical pion mass, as is summarized in the left pane of Figure \ref{fig:gafits}.
Alternatively we can fit with an additional finite-size effect term such as \(\exp(-m_\pi L)\) to the linear form and obtain values such as 1.20(6) as is described in the right pane of Figure \ref{fig:gafits}.
Detailed form of such finite-size effect term does not matter at the present level of calculations as our dynamic range in the scaling variable, \(m_\pi L\), is limited, yet the variance among them yield an estimate of the associated systematics: thus we quote a value \(g_A/g_V=1.20(6)_{\rm stat.}(4)_{\rm syst.}\).

\section{Momentum dependence of form factors}

Let us first discuss the vector current quantities, the vector and induced tensor form factors.
The vector charge, \(g_V=G_V(0)\), provides a check for the vector and axial current renormalization, \(Z_V=Z_A=1/g_V\).
Since the vector current is conserved, the vector charge is not affected by excited-state contamination, and hence is calculated very accurately.
The obtained value of \(Z_V=1/1.3918(19)=0.7185(10)\) in the chiral limit agrees very well with an independent calculation, \(Z_A=0.7161(1)\), in the meson sector \cite{Allton:2008pn} upto small \(O(m_fa)^2\) discretization error.

With the vector charge providing the normalization, we find the vector form factor, \(G_V (q^2)\), is fit well by the conventional dipole form, \(g_V/(1+\langle r_1^2\rangle q^2/12)\).
The single fitting parameter, the dipole coefficient, \(\langle r_1^2\rangle\), is the isovector Dirac mean-squared radius.
\begin{figure}[b]
\begin{center}
\includegraphics[width=.48\textwidth,clip]{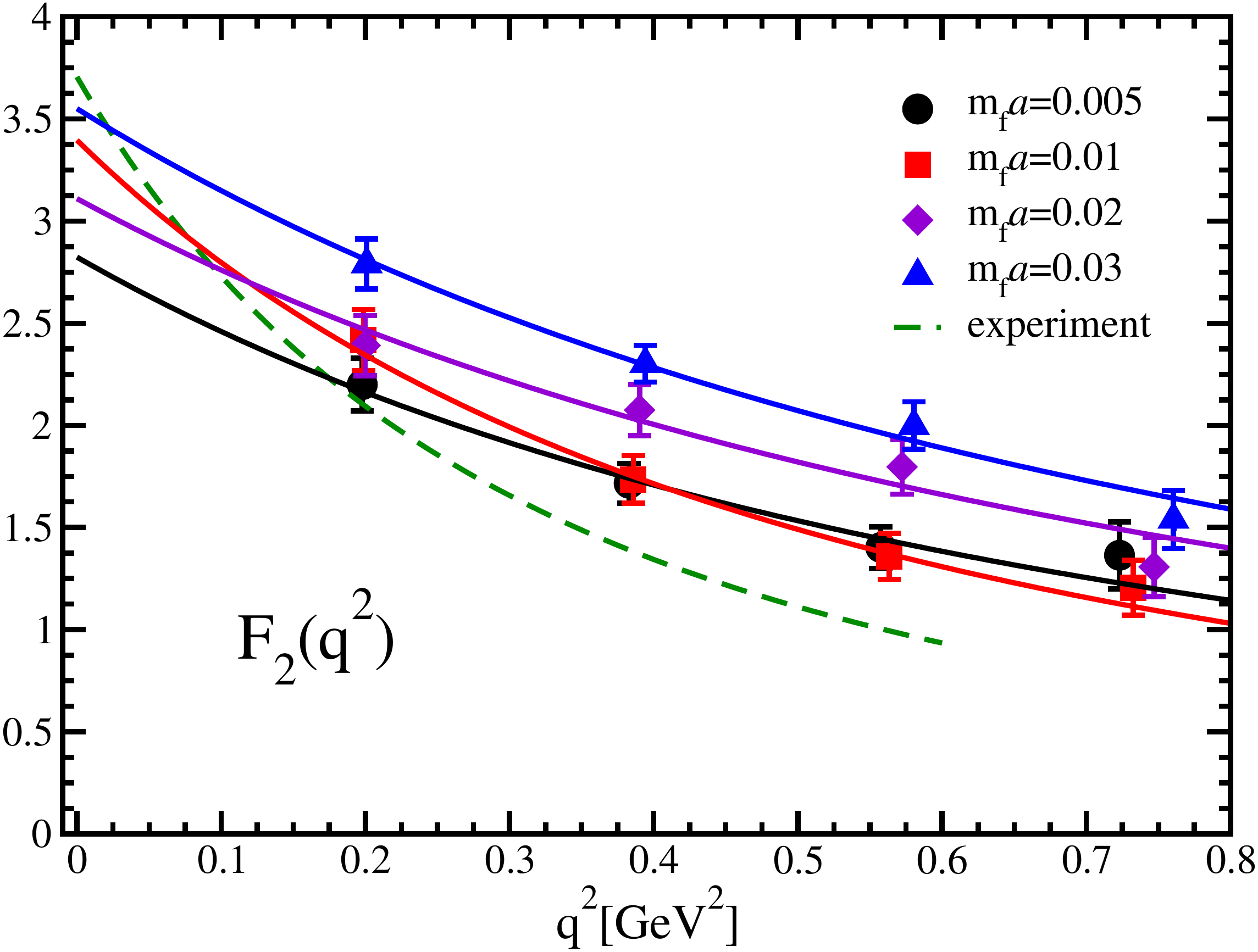}
\includegraphics[width=.48\textwidth,clip]{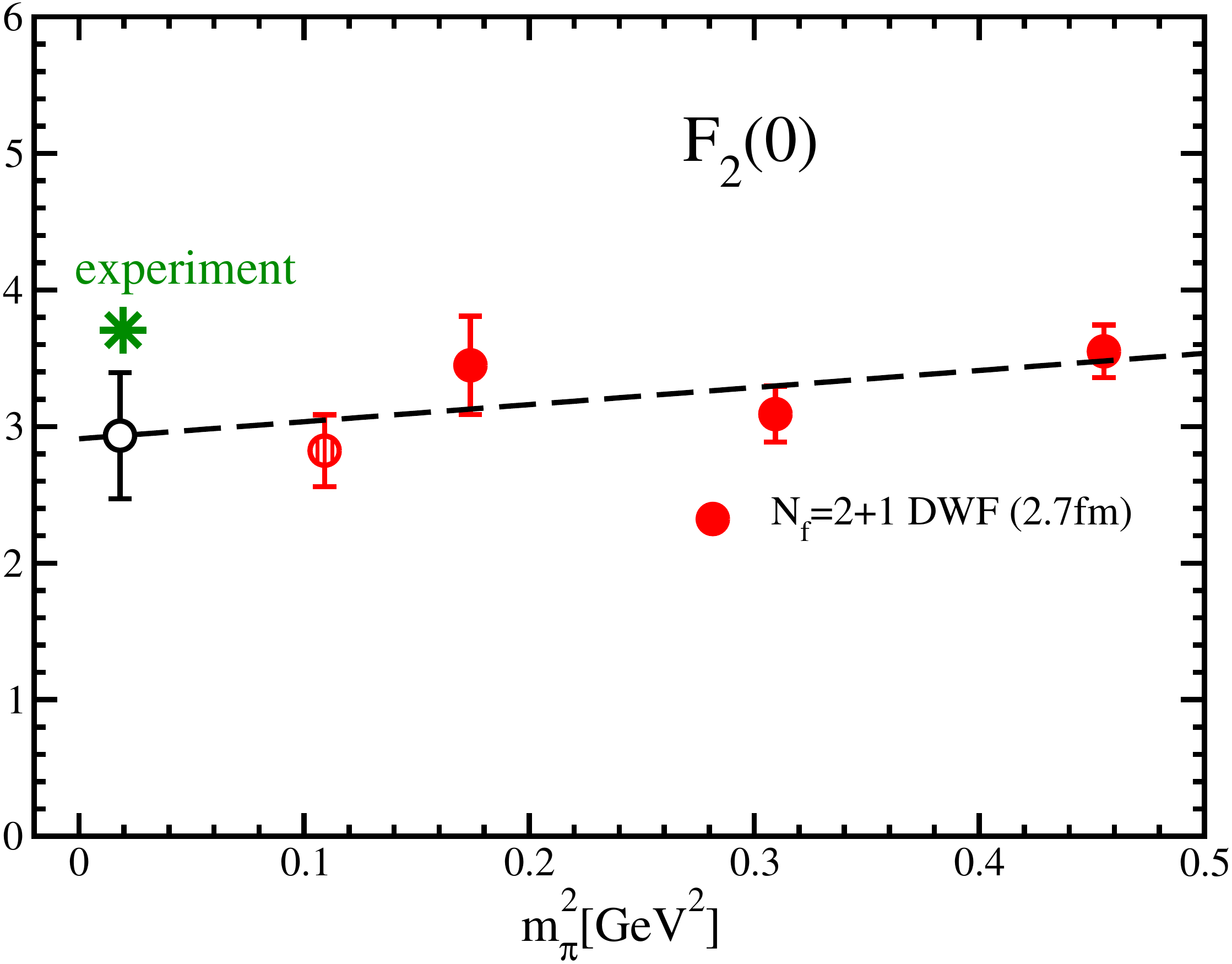}
\end{center}
\caption{Left: Dipole fit for the induced tensor form factor.  Right: Resultant anomalous magnetic moment, \(\mu_p-\mu_n-1\), of nucleon.  The linear fit exclude the lightest point (striped circle).  It extrapolates to within reasonable neighborhood of the experiment.\label{fig:GT}}
\end{figure}
\begin{figure}
\begin{center}
\includegraphics[width=.48\textwidth,clip]{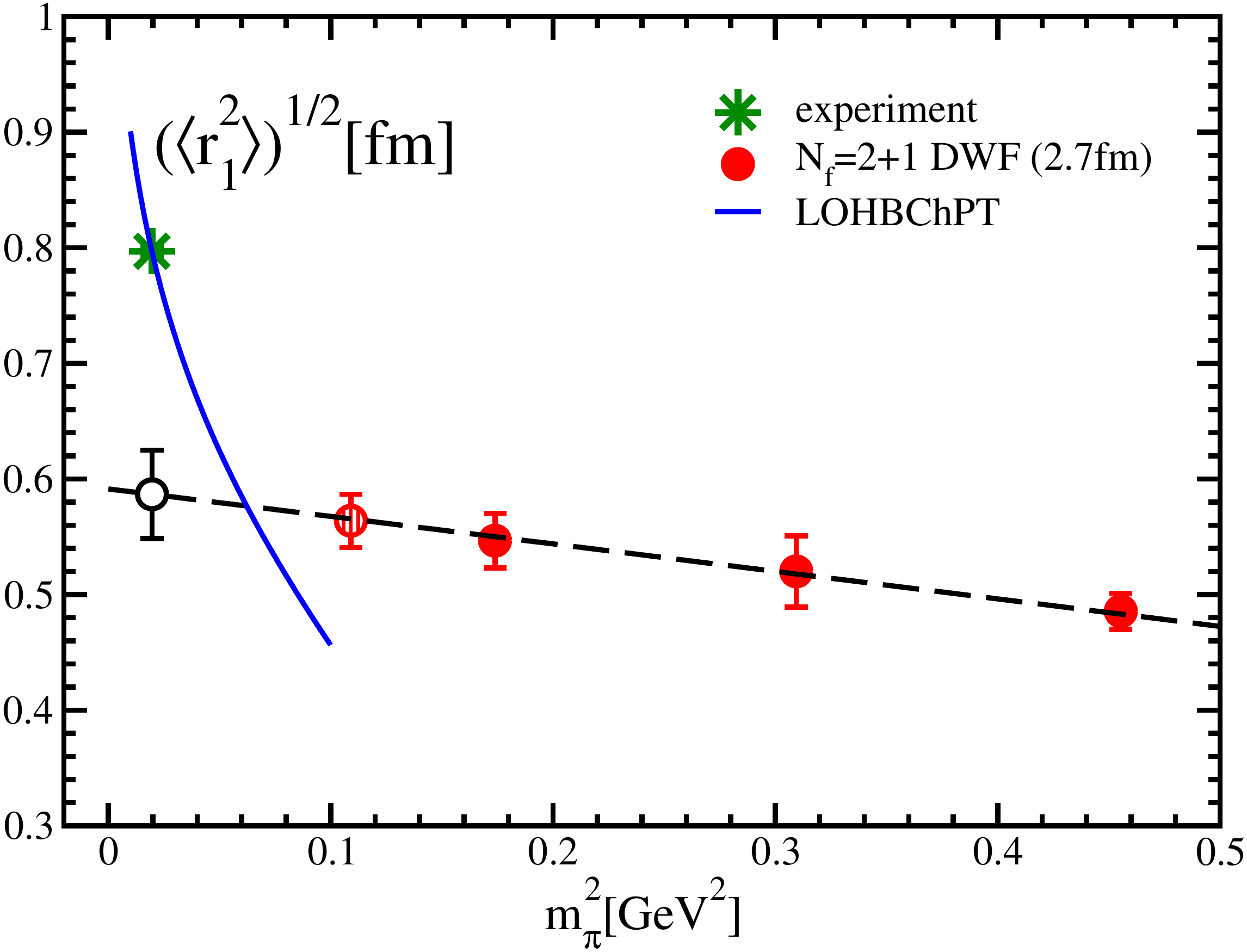}
\includegraphics[width=.48\textwidth,clip]{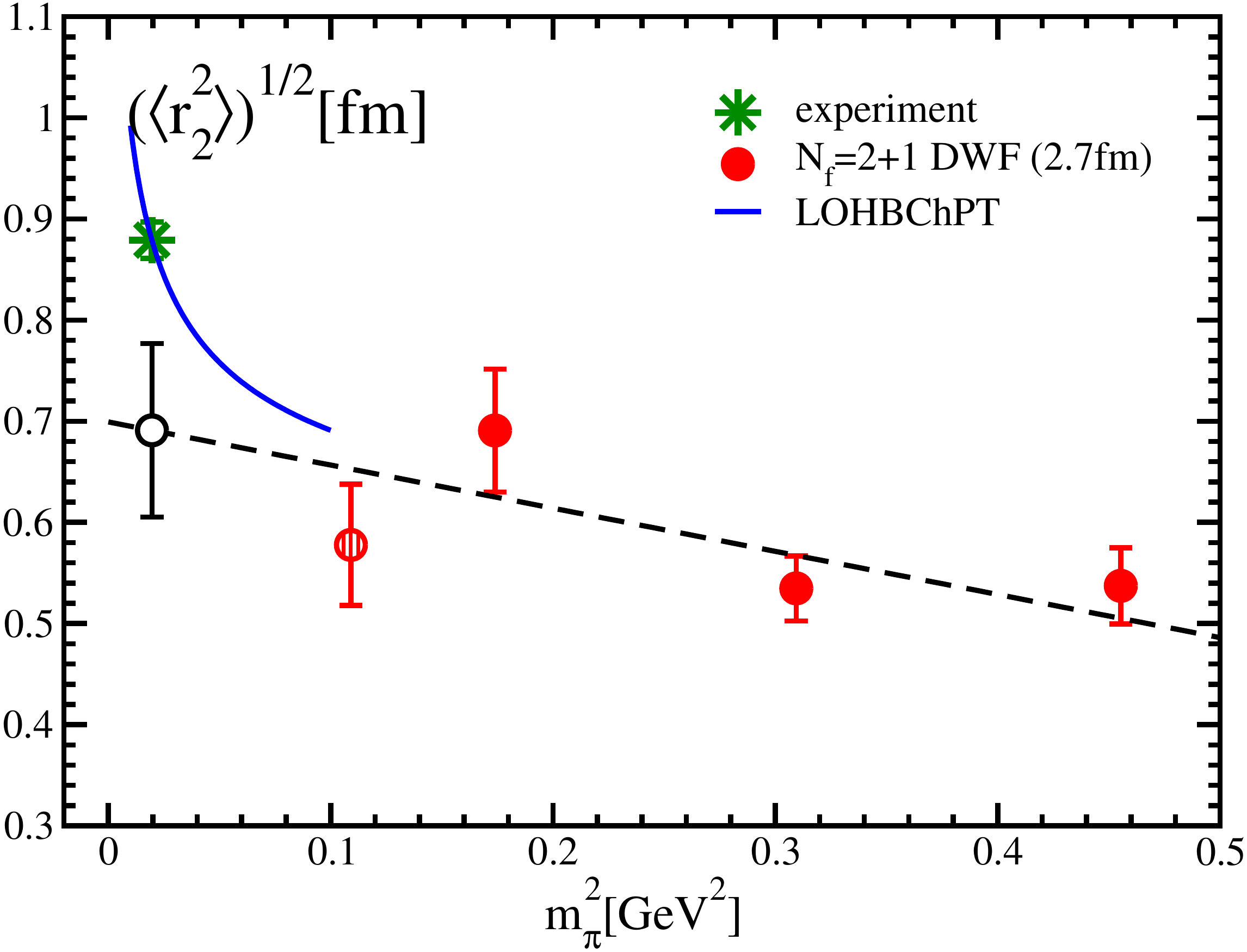}
\end{center}
\caption{Dirac and Pauli root mean-squared radii of nucleon.  Linear fits exclude the lightest points (striped circles).  Though they increase linearly toward the lighter physical pion mass, the expected divergent chiral behavior is yet to be seen.  Whether the lowest order heavy-baryon chiral perturbation model can describe such divergence is not clear in the Dirac radius, while it may still be consistent with the Pauli radius.\label{fig:DPradii}}
\end{figure}
The induced tensor form factor, \(G_T (q^2)\), is also fit well by the dipole form, \(F_2(0)/(2m_N)/(1+\langle r_2^2\rangle q^2/12)\), (see Figure \ref{fig:GT}.)
Here the strength, \(F_2(0)\), is a fitting parameter, and 
provides the isovector anomalous magnetic moment of nucleon, \(\mu_p-\mu_n-1\).
It agrees reasonably well with the experiment.
The second fitting parameter, the dipole coefficient, \(\langle r_2^2\rangle\), is the isovector Pauli mean-squared radius.

The dependence of the isovector Dirac, \(\langle r_1^2\rangle\), and Pauli, \(\langle r_2^2\rangle\), mean-squared radii on the up/down quark mass is summarized in Figure \ref{fig:DPradii}.
\begin{figure}[b]
\begin{center}
\includegraphics[width=.48\textwidth,clip]{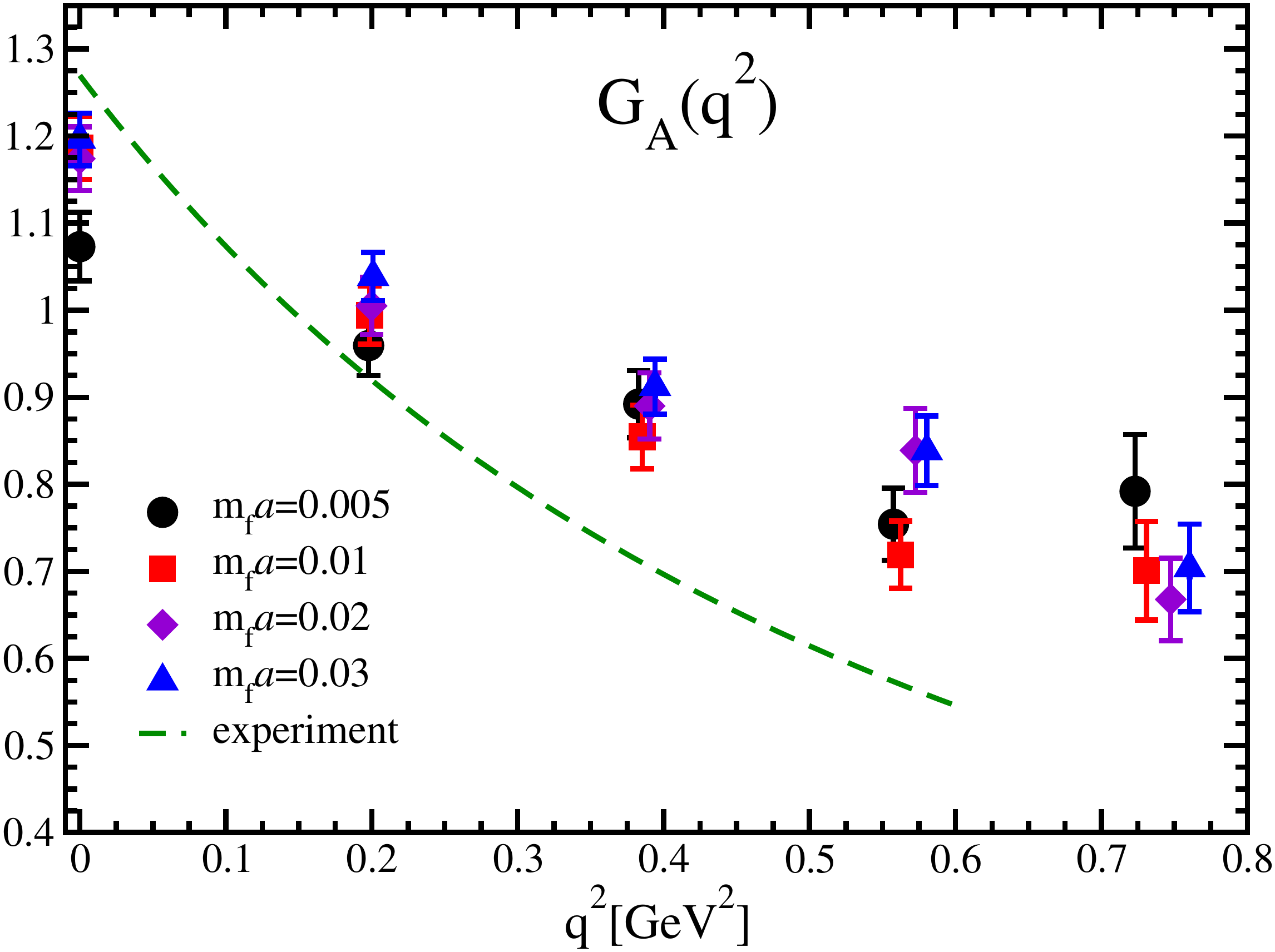}
\includegraphics[width=.48\textwidth,clip]{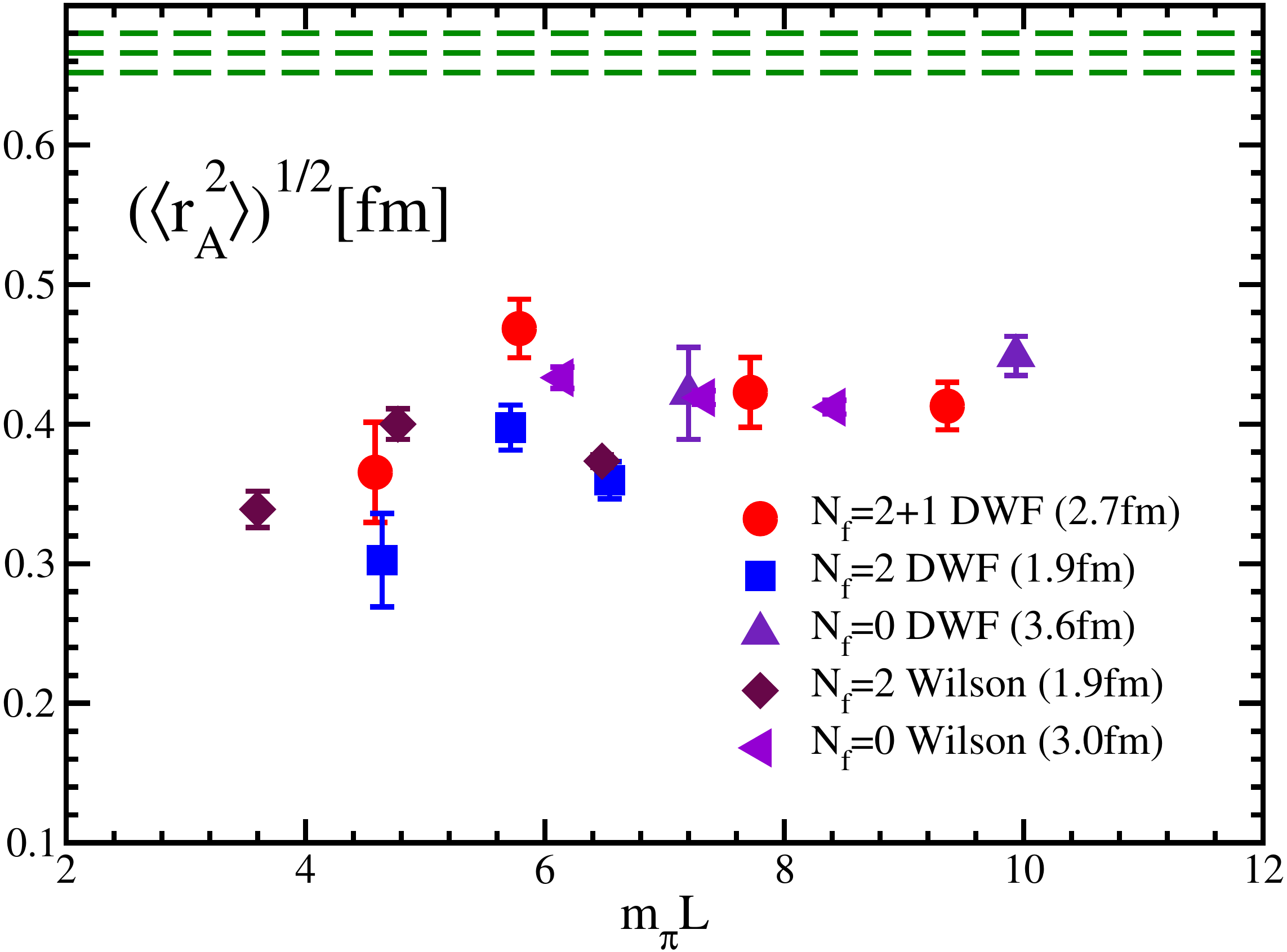}
\end{center}
\caption{Dipole fits to axialvector form factor (left) and resultant root mean-squared radius in comparison with other lattice calculations (right).  Green dashed lines indicate experiments.  The calculated mean-squared radius is well below the experiment but shows scaling in \(m_\pi L\) similar to the axial charge.\label{fig:aff}}
\end{figure}
They both increase toward decreasing pion mass and are easily fit by linear forms, yielding extrapolations to physical pion mass that undershoot experiments.
We do not see any sign of chirally expected divergent behavior in the chiral limit yet.
Whether any of the chiral-perturbation-inspired models can describe these mean-squared radii caculated on the lattice remains to be seen: the lowest-order "heavy-baryon chiral perturbation" seems inconsistent with our Dirac radius result but may be consistent with the Pauli one.

Let us now turn our attention to the axial-current form factors: 
The axialvector form factor admits single-parameter dipole fit (see the left pane of Figure \ref{fig:aff}).
The dipole coefficient, \(\langle r_A^2\rangle\), stays constant for large values of the scaling variable \(m_\pi L\) about 40 \% smaller than the experiment and independent of the number of dynamical flavors, and then deviates away further downward at \(m_\pi L \sim 5\) and smaller where only dynamical results are available.
This behavior corroborates the axial currents suffer from large finite-size effect.
Here at finite momentum we do not know yet if the effect is enhanced with dynamical quarks as there is no quenched calculation with light valence quarks with \(m_\pi L < 6\).

\begin{figure}
\begin{center}
\includegraphics[width=.48\textwidth,clip]{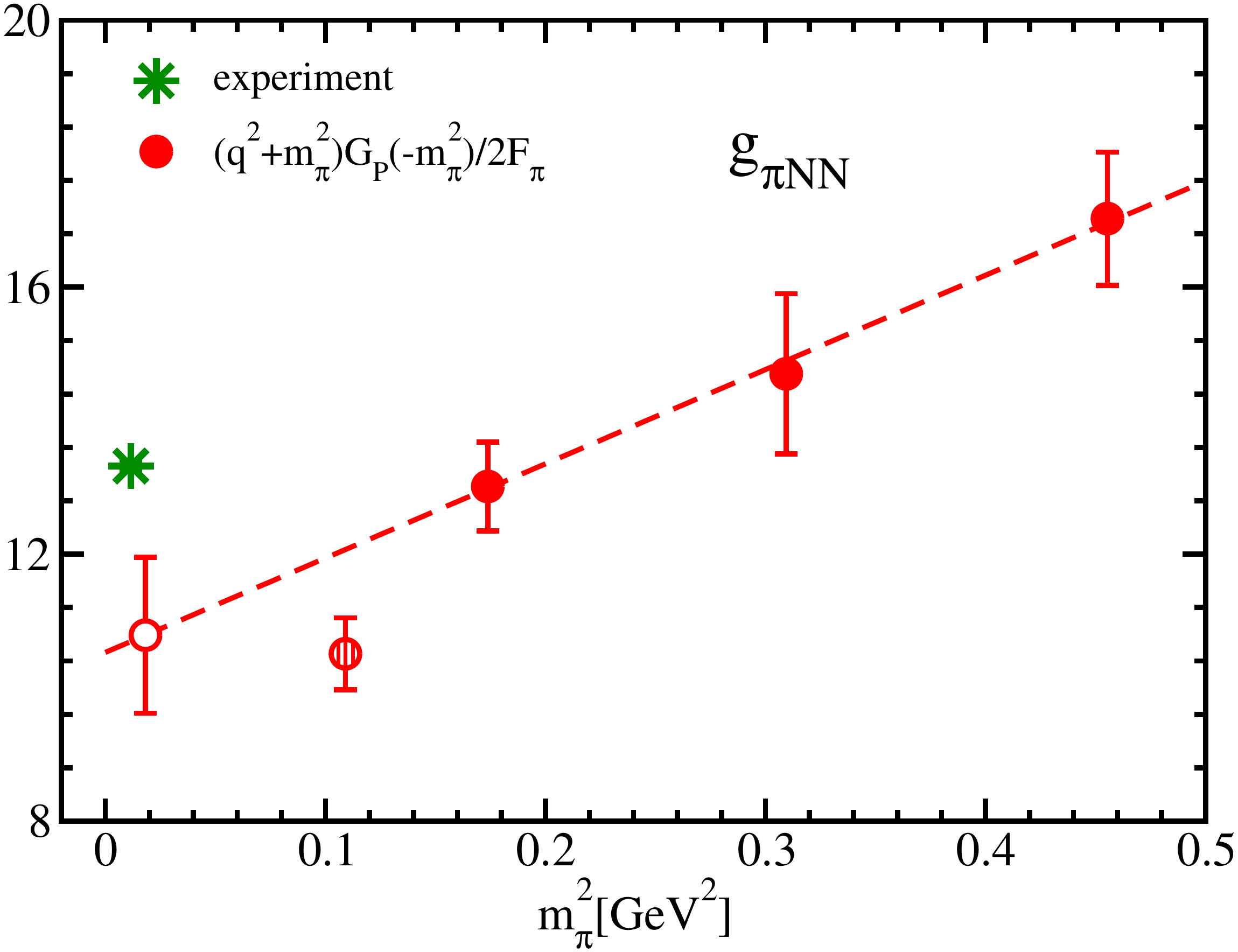}
\includegraphics[width=.48\textwidth,clip]{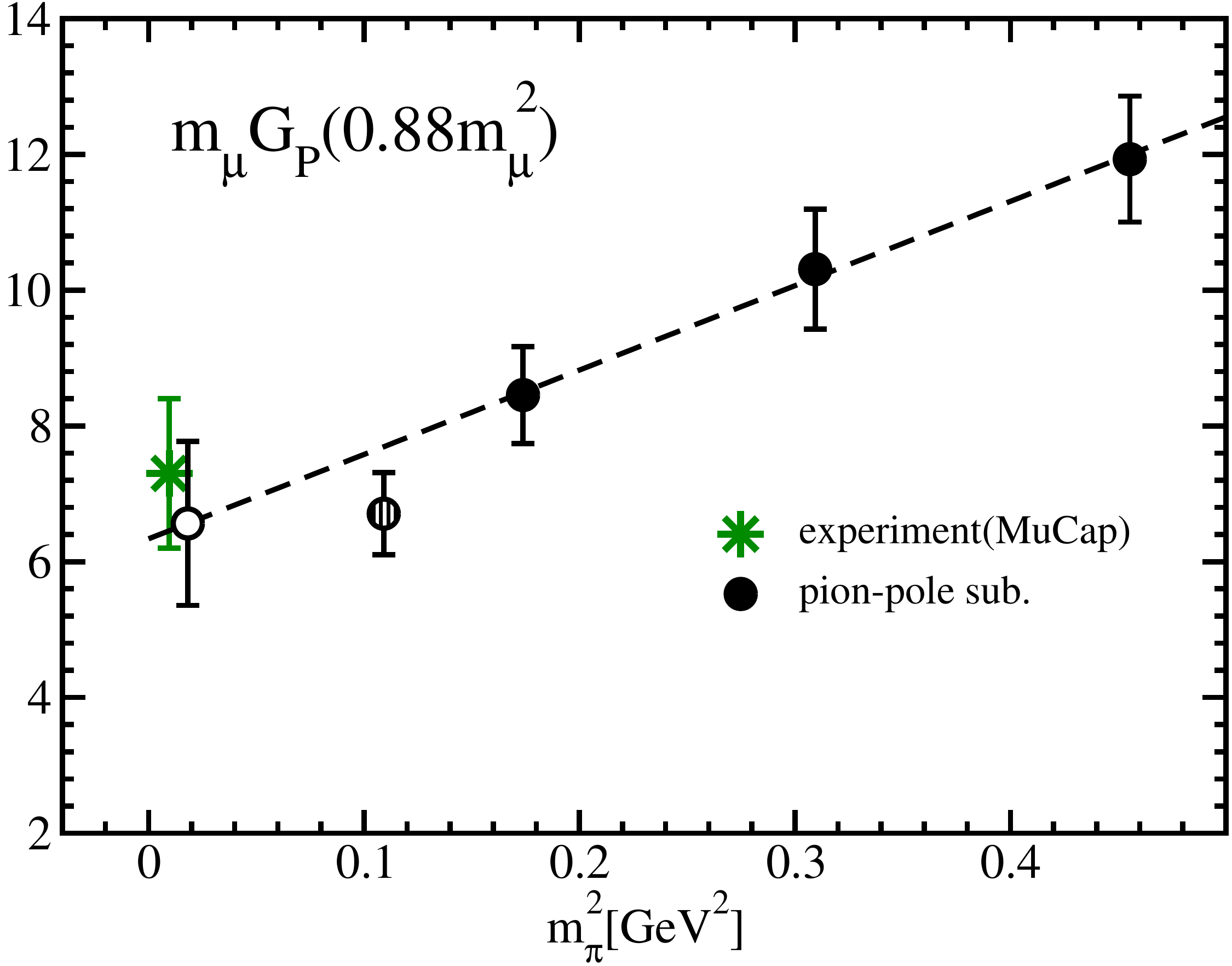}
\end{center}
\caption{Left: Pion-nucleon coupling, \(g_{\pi NN}\), extracted from near the pion pole, \(q^2\sim -m_\pi^2\) , of the calculated induced pseudoscalar form factor.  The linear fit excludes the deviating lightest point, and yet undershoots the experiment.  Right: muon-capture coupling \(g_P=m_\mu G_P(q^2=0.88 m_\mu^2)\), also extracted from the calculated induced pseudoscalar form factor.  The linear fit again exclude the slightly deviating lightest point which would not affect the agreement with the experiment too much.\label{fig:pff}}
\end{figure}

From the induced pseudoscalar form factor, \(G_P(q^2)\), we can extract the venerated pion-nucleon coupling, \(g_{\pi NN}\), from the residue at the pion pole, \(G_P(q^2) \sim 2 F_\pi g_{\pi NN}/ (q^2+m_\pi^2)\), \(q^2\sim -m_\pi^2\).
As is seen in the left pane of Figure \ref{fig:pff}, the results for the three heavier up/down quark mass linearly extrapolate well within two standard deviation of the experimental value.
The lightest point, however, shows downward shift away from the linear fit and the experiment, much like the axial charge.
Indeed this coupling, \(g_{\pi NN}\), and the axial charge, \(g_A\), are related by the Goldberger-Treiman relation \cite{Goldberger:1958tr}, \(\displaystyle
q^2 G_P(q^2) - 2 m_N G_A(q^2) = 
- \frac{g_{\pi NN} F_\pi m_\pi^2}{q^2+m_\pi^2}
\).

Another interesting experiment on the induced pseudoscalar form factor is muon capture.
We extract from our calculation the effective coupling, \(g_P=m_\mu G_P(q^2=0.88 m_\mu^2)\), for this process.
As is shwon in the right pane of Figure \ref{fig:pff}, our results compare favorably with the recent MuCap result \cite{Andreev:2007wg}.
Here again we see the lightest point deviating downward, away from the linear fit to the three heavier points and the experiment.
But the effect is less significant and does not affect the seemingly good agreement with the experiment.

\section{Moments of structure functions}

\begin{figure}
\begin{center}
\includegraphics[width=.48\textwidth,clip]{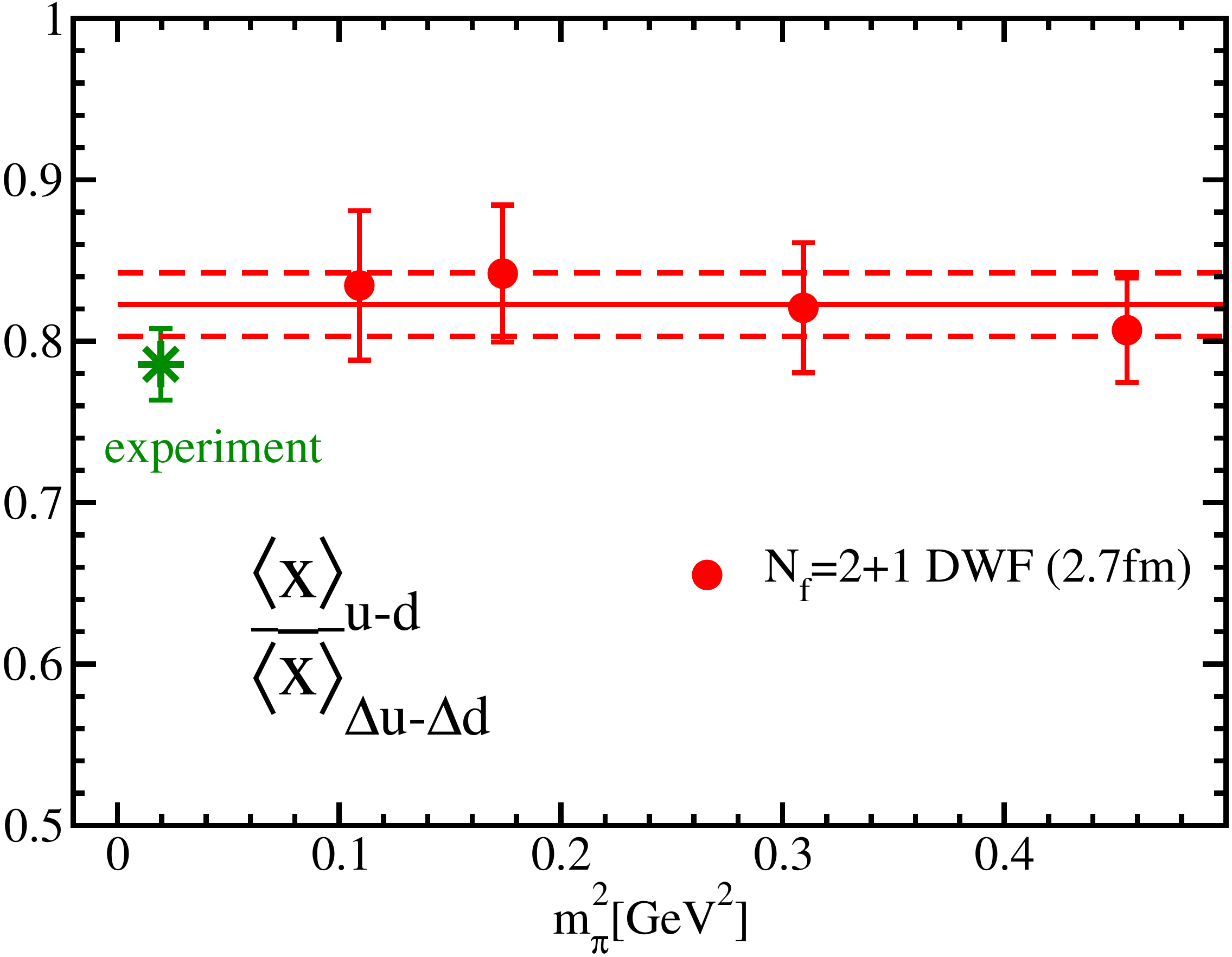}
\end{center}
\caption{Naturally renormalized ratio, \(\langle x \rangle_{u-d}/\langle x \rangle_{\Delta u - \Delta d}\), of the isovector quark momentum fraction to the helicity fraction.\label{fig:xqxDqratio}}
\end{figure}

Let us first discuss the lattice ratio, \(\langle x \rangle_{u-d}/\langle x \rangle_{\Delta u - \Delta d}\), of the isovector quark momentum fraction to the helicity fraction.
Much like the form factor ratio, \(g_A/g_V\), the momentum fraction, \(\langle x \rangle_{u-d}\), the first moment of the \(F_{1,2}\) unpolarized structure functions, and helicity fraction, \(\langle x \rangle_{\Delta u - \Delta d}\), the first moment of the \(g_1\) polarized structure function, share a common renormalization as they are related by a chiral rotation and the DWF preserve the chiral symmetry.
Thus the ratio calculated on the lattice is directly comparable with the experiment.
The results of our calculation are shown in Figure \ref{fig:xqxDqratio}.
They do not show any discernible dependence on the up/down quark mass, and are in excellent agreement with the experiment.
This is in contrast to the similarly naturally renormalized ratio of \(g_A/g_V\) of elastic form factors which at the lightest point deviates significantly from heavier mass results and the experiment.
This suggests the moments of inelastic structure functions such as the momentum fraction, \(\langle x \rangle_{u-d}\), and helicity fraction, \(\langle x \rangle_{\Delta u - \Delta d}\), may not suffer so severely from the finite-size effect that plagues elastic form factor calculations.
To clarify this, we are performing follow up calculations at the smaller volume of 1.8 fm.

\begin{figure}
\begin{center}
\includegraphics[width=.48\textwidth,clip]{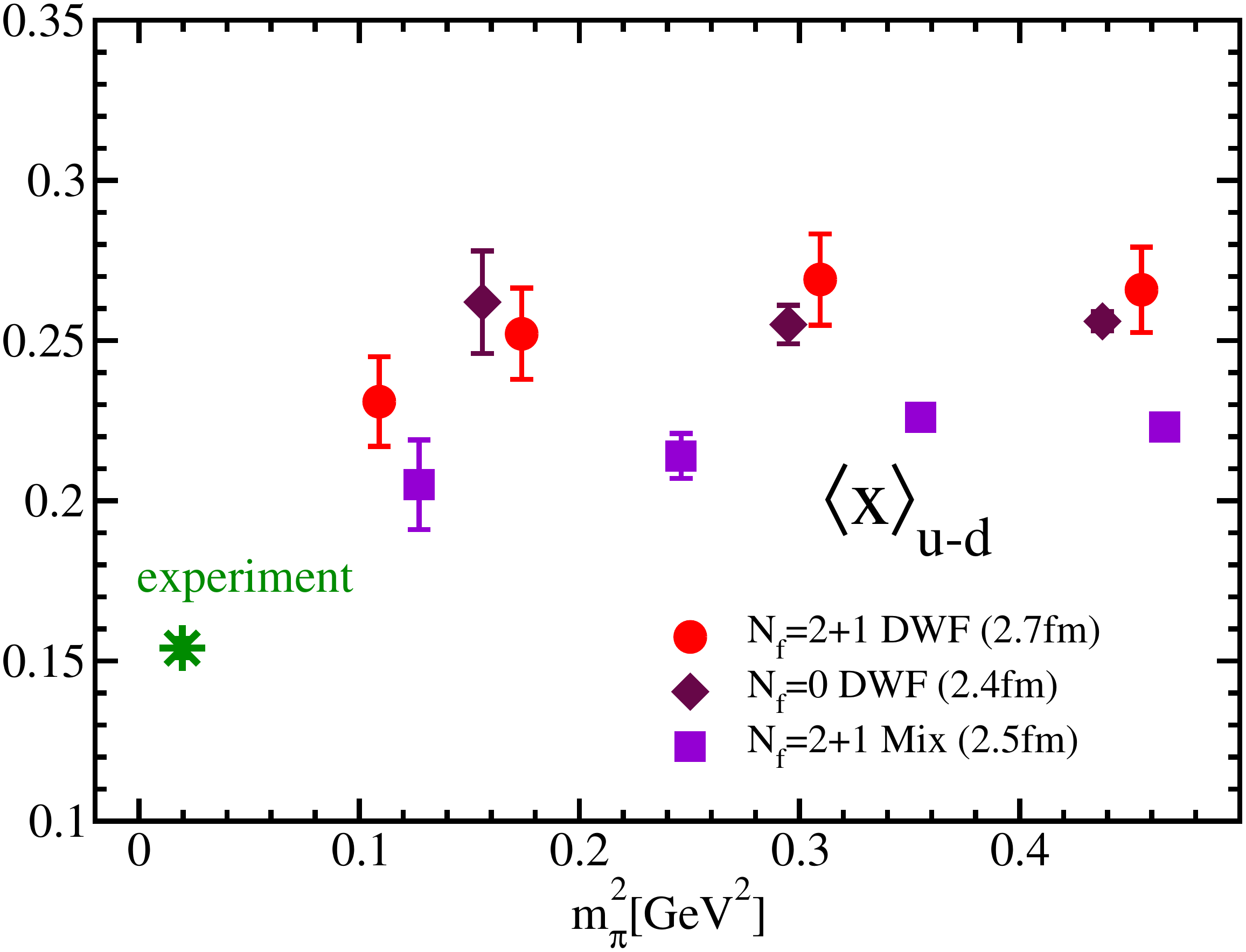}
\includegraphics[width=.48\textwidth,clip]{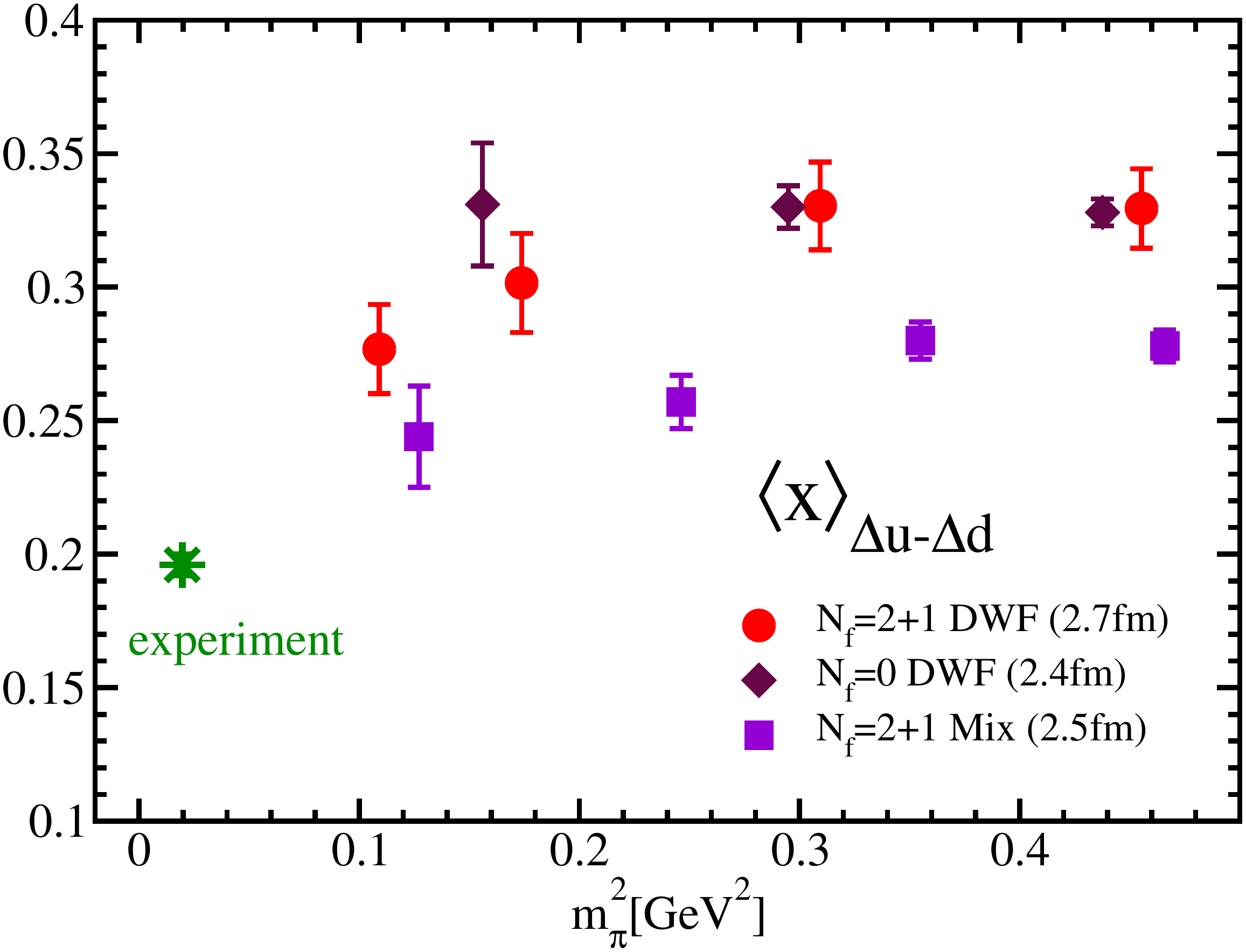}
\end{center}
\caption{Absolute value of the quark momentum fraction, \(\langle x \rangle_{u-d}\), fully non-perturbatively renormalized (left), and the helicity fraction, \(\langle x \rangle_{\Delta u - \Delta d}\), also fullly non-perturbatively renormalized (right).\label{fig:xqxDq}}
\end{figure}

Next we discuss the absolute values of the isovector quark momentum fraction, \(\langle x \rangle_{u-d}\).
This is the first moment of the unpolarized structure functions, \(F_1\) and \(F_2\).
It is now fully non-perturbatively renormalized and run to \(\overline{\rm MS}\) at 2 GeV, \(Z^{\overline{\rm MS}}({\rm 2 GeV}) = 1.15(4)\).
The results are shown in the left pane of Figure \ref{fig:xqxDq}.
On the larger, 2.7 fm, volume, the three heavier points stay roughly the constant which is about 70 \% higher at \(\sim 0.26\) than the experiment, about 0.15.
This behavior is not so different from old RBC quenched results \cite{Orginos:2005uy} wtih similar up/down quark mass.
The lightest point shows a sign of deviation away from this heavy constant behavior, but in contrast to the form factor deviations, is trending toward the experiment.
Since lighter quark can more easily share its momentum with other degrees of freedom, this trending toward the experiment may well be a real physical effect:
It is not necessarily be affected by the finite spatial size of the lattice.
To clarify,  we are performing follow up calculations at \(m_fa=0.01\) on the smaller volume of 1.8 fm, with \(m_\pi L\) smaller than the \(m_fa=0.005\) on 2.7 fm volume.
If this follow up calculation stays with the higher value, \(\sim 0.26\), then this moment is not so seriously affected by the finite-size effect.

In the Figure we also compare our results with the LHPC/MILC mixed action calculations \cite{Hagler:2007xi}.
Their values are significantly lower, by about 20 \%, than ours.
The difference is likely either caused by their use of only perturbative renormalization or the significantly shorter source/sink separation in time, or both.
Another possibility is the non-unitary matching of the DWF and rooted-staggered fermions.

The isovector quark helicity fraction, \(\langle x \rangle_{\Delta u - \Delta d}\), is also fully non-perturbatively renormalized and run to \(\overline{\rm MS}\) at 2 GeV.
The renormalization agrees very well with the momentum fraction one, \(Z^{\overline{\rm MS}}({\rm 2 GeV}) = 1.15(3)\).
As is shown in the right pane of Figure \ref{fig:xqxDq}, the observable exhibits very similar behavior to the momentum fraction, as can be expected from the near constancy of their ratio: the three heavier points stay roughly the constant which is about 70 \% higher than the experiment.
This is again not so different from old RBC quenched results \cite{Orginos:2005uy} wtih similar up/down quark mass.
The lightest point then again shows a sign of deviation away from this heavy constant behavior, and again in contrast to the form factor deviations, is trending toward the experiment.
Again this trending toward the experiment may well be a real physical effect, or it can be affected by the finite spatial size of the lattice.
The situation may be clarified by the on-going small-volume calculations.

Again the LHPC/MILC mixed action results are significantly lower than ours.
And again the difference is likely either caused by their use of only perturbative renormalization or the significantly shorter source/sink separation in time, or their combination.
Yet again another possibility is to the non-unitary matching of the DWF and rooted-staggered fermions.

\begin{figure}
\begin{center}
\includegraphics[width=.48\textwidth,clip]{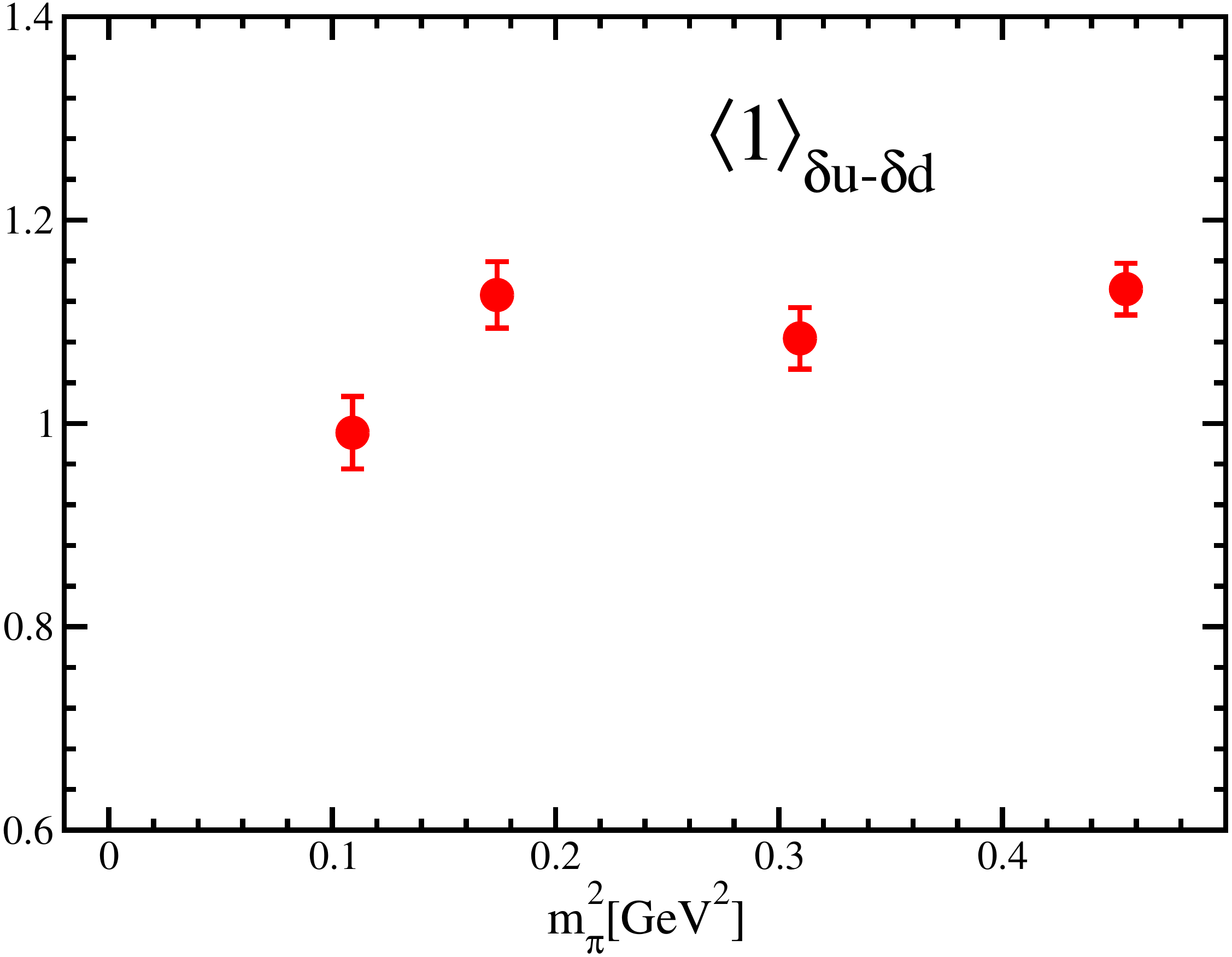}
\includegraphics[width=.48\textwidth,clip]{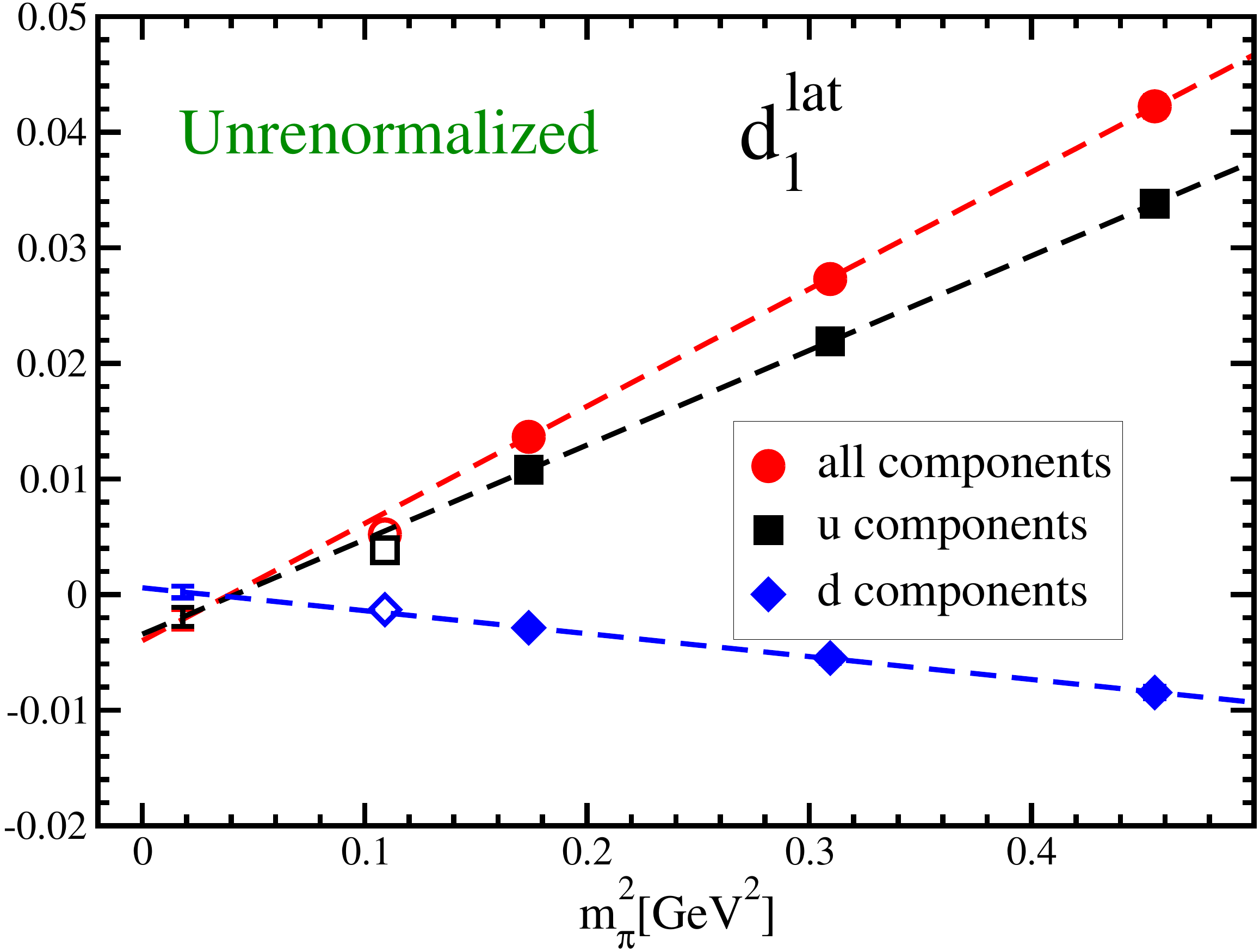}
\end{center}
\caption{Absolute value of the tensor charge, \(\langle 1 \rangle_{\delta u-\delta d}\), fully non-perturbatively renormalized (left), and the twist-3, \(d_1\), moment of the \(g_2\) structure function, not renormalized (right).\label{fig:1qd1}}
\end{figure}

The results for the isovector tensor charge, \(\langle 1 \rangle_{\delta u - \delta d}\), are presented in the left pane of Figure \ref{fig:1qd1}.
These are fully non-perturbatively renormalized and run to \(\overline{\rm MS}\) at 2 GeV, with \(Z^{\overline{\rm MS}}({\rm 2 GeV}) = 0.783(3)\).
These can provide a prediction since the experiments are yet to report a number.
If we fit the heavy three points with a constant we obtain a value of about 1.10(7).
Alternatively if we linearly extrapolate the two lightest points we would obtain about 0.7.
Since we do not understand the finite-size correction nor chiral extrapolation, these values are about as good as we can make.

In the right pane of Figure \ref{fig:1qd1} we present the twist-3 moment, \(d_1\), of the polarized structure function \(g_2\).
Unlike the other structure functions we discussed in the above, this one is not renormalized yet.
Our interest here is in whether the perturbatively obtained Wandzura-Wilczek relation \cite{Wandzura:1977qf} holds.
From the smallness of the values obtained, we conclude it holds.
We note we have the lightest mass points deviating from the linear trends set by the heavier points.

\section{Conclusions}

We calculated all the four nucleon isovector form factors, vector, induced tensor, axialvector and induced pseudoscalar, and some low moments of isovector nucleon structure functions on the lattice with (2+1)-flavor dynamical DWF quarks.
The lattice cutoff is about 1.7 GeV, and two spatial volumes, one 2.7 fm across and the other 1.8 fm, are investigated.
The dynamical strange quark mass is set about 12 \% heavier than physical, and the degenerate up/down quark mass is varied corresponding to pion mass of about 0.67, 0.56, 0.42 and 0.33 GeV and nucleon mass about 1.55, 1.39, 1.22 and 1.15 GeV.
We carefully chose the nucleon source and sink separation in time as long as possible, at about 1.4 fm, in order to eliminate excited-state contamination as much as possible while maintaining the signal.
We suspect anything shorter would result in unacceptable contamination for the observables.

A big finite-size effect is seen at the light up/down quark mass in the axial charge, \(g_A/g_V\), a naturally renormalized observable with DWF quarks that preserve the chiral symmetry.
The effect takes this observable away from the experimental value.
It scales with a single variable, \(m_\pi L\), and sets in at about \(m_\pi L \sim 5\).
Similar effects are seen in axialvector and induced pseudoscalar form factors, but are absent in vector and induced tensor form factors:
This is because the axial current is not conserved while the vector current is.
Small volume calculations no longer make sense, at least for axialvector and induced pseudoscalar form factors.
We plan to use auxiliary determinant to make larger volume calculations possible.
The induced tensor form factor yields a reasonable value for the isovector anomalous magnetic moment of nucleon, \(\mu_p - \mu_n -1\), when linearly extrapolated to physical pion mass.
The Dirac and Pauli isovector mean-squared radii increase toward lighter, physical pion mass as well, but does not show any sign of chiral divergence.

Though similarly naturally renormalized, the ratio, \(\langle x \rangle_{u-d}/\langle x \rangle_{\Delta u - \Delta d}\), of the quark momentum and helicity fractions, is not affected by such a finite-size effect.
It does not depend on light quark mass either and agree well with the experiment.
Thus the calculations of the moments of inelasitc structure functions may still make sense on small volumes.
Their respective absolute values, fully non-perturbatively renormalized, shows interesting trending toward the respective experimental values at the lightest light quark mass.
This may well be real physical effect driven by lighter quarks, in contrast to the finite-size effect in the axial-current form factors.
Our on-going follow-up calculations at \(m_fa=0.01\) on the smaller volume of 1.8 fm, with \(m_\pi L\) smaller than the \(m_fa=0.005\) on 2.7 fm volume, has a potential to resolve this question.

In addition, the fully non-perturbatively renormalized tensor charge, \(\langle 1 \rangle_{\delta u - \delta d}\) has been obtained.
We need to understand the finite-size corrections or chiral extrapolations before transforming the values into a prediction for the upcoming experiments. 

The twist-3, \(d_1\), moment of the \(g_2\) structure function is also obtained.
Though yet to be renormalized, its smallness suggests the Wandzura-Wilczek relation holds.

We thank the members of the RBC and UKQCD Collaborations, in particular, Yasumichi Aoki, Tom Blum, Huey-Wen Lin, Meifeng Lin, Shoichi Sasaki, Robert Tweedie and James Zanotti.
We also thank the RIKEN-BNL Research Center for partial support.
TY is the Yukawa Fellow supported by Yukawa Memorial Foundation.
RIKEN, BNL, the U.S. DOE, Edinburgh University and the UK PPARC provided facilities essential for this work.

\end{document}